\documentclass[fleqn,usenatbib]{mnras}

\usepackage{newtxtext,newtxmath}

\usepackage[T1]{fontenc}

\DeclareRobustCommand{\VAN}[3]{#2}
\let\VANthebibliography\thebibliography
\def\thebibliography{\DeclareRobustCommand{\VAN}[3]{##3}\VANthebibliography}

\usepackage{graphicx}	
\usepackage{amsmath}	

\usepackage{amssymb}	

\title[WALLABY Pilot Survey: RPS of ESO\,501$-$G075]{WALLABY Pilot Survey: First Look at the Hydra I Cluster and Ram Pressure Stripping of ESO\,501$-$G075}

\author[T.N.~Reynolds, et al.]{T.N.~Reynolds,$^{1,2}$\thanks{tristan.reynolds@uwa.edu.au} 
T.~Westmeier,$^{1,2}$
A.~Elagali,$^{3,2,13}$
B.~Catinella,$^{1,2}$
L.~Cortese,$^{1,2}$
N.~Deg,$^{10}$
B.-Q.~For,$^{1,2}$
\newauthor
P.~Kamphuis,$^{11}$
D.~Kleiner,$^{5}$
B.S.~Koribalski,$^{4,6}$
K.~Lee-Waddell,$^{1,3}$
S.-H.~Oh,$^{7}$
J.~Rhee,$^{1,2}$
P.~Serra,$^{5}$
\newauthor
K.~Spekkens,$^{9}$
L.~Staveley-Smith,$^{1,2}$
A.R.H.~Stevens,$^{1,2}$
E.N.~Taylor,$^{12}$
J.~Wang,$^{8}$
O.I.~Wong$^{3,1,2}$
\\
$^1$International Centre for Radio Astronomy Research (ICRAR), The University of Western Australia, 35 Stirling Hwy, Crawley, WA, 6009, Australia\\
$^2$ARC Centre of Excellence for All Sky Astrophysics in 3 Dimensions (ASTRO 3D) \\
$^3$CSIRO Astronomy and Space Science, PO Box 1130, Bentley WA 6102, Australia \\
$^4$CSIRO Astronomy and Space Science, Australia Telescope National Facility, P.O. Box 76, Epping NSW 1710, Australia \\
$^5$INAF – Osservatorio Astronomico di Cagliari, Via della Scienza 5, 09047 Selargius, CA, Italy \\
$^6$Western Sydney University, Locked Bag 1797, Penrith, NSW 2751, Australia \\
$^7$Department of Physics and Astronomy, Sejong University, 209 Neungdong-ro, Gwangjin-gu, Seoul, Republic of Korea \\
$^8$Kavli Institute for Astronomy and Astrophysics, Peking University, Beijing 100871, China \\
$^9$Department of Physics and Space Science Royal Military College of Canada P.O. Box 17000, Station Forces Kingston, ON K7K 7B4, Canada \\
$^{10}$Department of Physics, Engineering Physics, and Astronomy, Queen's University, Kingston, ON, K7L 3N6, Canada \\
$^{11}$Ruhr-Universität Bochum, Faculty of Physics and Astronomy, Astronomical Institute, 44780 Bochum, Germany \\
$^{12}$Centre for Astrophysics and Supercomputing, Swinburne University of Technology, John Street, Hawthorn VIC 3122, Australia \\
$^{13}$Telethon Kids Institute, Perth Children’s Hospital, Perth, Australia
}

\date{Accepted 2021 May 10. Received 2021 April 28; in original form 2020 October 26}

\pubyear{2021}

\begin{document}
\label{firstpage}
\pagerange{\pageref{firstpage}--\pageref{lastpage}}
\maketitle

\begin{abstract}
We present results from neutral atomic hydrogen (H\,\textsc{i}) observations of Hydra I, the first cluster observed by the Widefield ASKAP L-band Legacy All-sky Blind Survey (WALLABY) on the Australian Square Kilometre Array Pathfinder. For the first time we show that WALLABY can reach its final survey sensitivity. Leveraging the sensitivity, spatial resolution and wide field of view of WALLABY, we identify a galaxy, ESO\,501$-$G075, that lies near the virial radius of Hydra I and displays an H\,\textsc{i} tail. ESO\,501$-$G075 shows a similar level of morphological asymmetry as another cluster member, which lies near the cluster centre  and shows signs of experiencing ram pressure. We investigate possible environmental processes that could be responsible for producing the observed disturbance in the H\,\textsc{i} morphology of ESO\,501$-$G075. We rule out tidal interactions, as ESO\,501$-$G075 has no nearby neighbours within $\sim0.34$\,Mpc. We use a simple model to determine that ram pressure can remove gas from the disc at radii $r\gtrsim25$\,kpc. We conclude that, as ESO\,501$-$G075 has a typical H\,\textsc{i} mass compared to similar galaxies in the field and its morphology is compatible with a ram pressure scenario, ESO\,501$-$G075 is likely recently infalling into the cluster and in the early stages of experiencing ram pressure.
\end{abstract}




\section{Introduction}
\label{sec:intro}

A galaxy's morphology is strongly influenced by the environment in which the galaxy resides. Late-type, spiral and irregular galaxies are more commonly found in lower density environments (isolated, in the field, low density groups and near the edges of clusters). The fraction of early-type, ellipticals increases at higher densities (towards the centre of clusters). This is well known as the morphology-density relation \citep[e.g.][]{Hubble1931,Oemler1974,Dressler1980}. The dramatic shift from large fractions of late-type galaxies at low densities through to low fractions near cluster centres \citep[$\sim60\%$ to $\sim5\%$, respectively, e.g.][]{Dressler1980,Houghton2015} indicates that significant morphological evolution occurs within galaxy groups and clusters. 

There are a number of environmental processes that affect the observed stellar and/or gas morphology of galaxies, including mergers \citep[e.g.][]{Toomre1972, Mihos1996, Zaritsky1997, Bournaud2005a, Elagali2018}, interactions between galaxies with low relative velocities \citep[tidal stripping, e.g.][]{Moore1999, Koribalski2009, English2010} and high relative velocities \citep[`harassment', e.g.][]{Moore1996,Moore1998}, ram pressure stripping by the intergalactic medium \citep{Gunn1972}, viscous stripping by the hot intergalactic medium in clusters \citep{Nulsen1982, Quilis2000, Rasmussen2006} and tidal stripping by a cluster's tidal field \citep[e.g.][]{Merritt1984, Gnedin2003a, Gnedin2003b}. Galaxy clusters are characterised by high densities of galaxies \citep[e.g.][]{Dressler1980}, large velocity dispersions \citep[e.g.][]{Girardi1993, Sohn2017} and a high density intergalactic medium (IGM) of ionized gas \citep[e.g.\ as measured from X-ray observations,][]{McDonald2017}. Observations and simulations find galaxies in clusters to be experiencing various combinations of these processes \citep[e.g.][]{Gunn1977, Moore1996, Moore1998, Gnedin2003a, Gnedin2003b, Kenney2004, Chung2007, Vollmer2009, Scott2010, Abramson2011, Bialas2015, Chen2020, Ramatsoku2020}. Of these processes, the proposed dominant driver for evolving galaxies' neutral atomic hydrogen (H\,\textsc{i}) morphology and quenching star formation is ram pressure stripping \citep[e.g.][]{Boselli2006}. Ram pressure stripping is also proposed to be the dominant mechanism responsible for observed H\,\textsc{i} deficiencies in cluster galaxies \citep[e.g.\ in the Virgo cluster,][]{Kenney2004, Chung2007, Yoon2017}.

These environmental processes can affect both the stellar and gaseous components of galaxies. Disentangling the relative importance that these processes play in the evolution of galaxy morphology and composition requires spatially resolved observations of the different galaxy components (stellar and/or gaseous). As a result, these environmental mechanisms are often first observed in the H\,\textsc{i} gas \citep[e.g.][]{Giovanelli1985, Solanes2001, Westmeier2011, Rasmussen2012}. H\,\textsc{i} observed beyond the edge of the optical disc resides further from the centre of the galaxy's potential well and can be more easily disturbed than the stellar disc. This makes H\,\textsc{i} an effective probe for investigating the influence of the environment.

There have been numerous studies of the H\,\textsc{i} content of galaxies in clusters in the local Universe using observations from both single-dish radio telescopes \citep[e.g.][]{Giovanelli1985, Barnes1997, Waugh2000, Solanes2001, Taylor2013, Scott2018} and radio interferometers \citep[e.g.][]{Cayatte1990, Cayatte1994, McMahon1993, Dickey1997, BravoAlfaro2000, BravoAlfaro2001, Verheijen2001, Chung2009}. These observations used parabolic dish antennas fitted with single pixel feed-horn receivers. Unlike observations taken with single dish telescopes, which are typically limited to measuring integrated galaxy properties, interferometric observations can spatially resolve galaxies, but are limited by their small instantaneous field of view. Hence, a spatially resolved survey of a targeted galaxy sample within a cluster requires significant integration time \citep[e.g.\ the VIVA\footnote{VLA Imaging of Virgo spirals in Atomic gas.} survey which observed 53 late-type galaxies in the Virgo cluster for $\sim8$\,h each,][]{Chung2009}. Surveys with instruments like the CSIRO Australian Square Kilometre Array Pathfinder \citep[ASKAP,][]{Johnston2008,Hotan2021} will help to fill this niche.

\subsection{The Australian Square Kilometre Array Pathfinder}
\label{s-sec:askap}

ASKAP is a new interferometer composed of 36 dish antennas fitted with phased array feed \citep[PAF,][]{DeBoer2009,Hampson2012,Hotan2014,Schinckel2016} receivers. The ASKAP PAFs consist of 188 dipole elements in a chequerboard pattern which can simultaneously form 36 beams on the sky, each with a primary beam size of $1^{\circ}$. The 30-square-degree field of view of ASKAP enables it to map the H\,\textsc{i} emission at an angular resolution of $\sim30$\,arcsec over large areas of sky more quickly than possible with other interferometric arrays (e.g.\ the Australia Telescope Compact Array or the Karl G. Jansky Very Large Array with a field of view of $\sim42$\,arcmin and $\sim32$\,arcmin at 1.4\,GHz, respectively).

The Widefield ASKAP L-band Legacy All-sky Blind Survey \citep[WALLABY,][]{Koribalski2020} is the H\,\textsc{i} all-sky survey being conducted on ASKAP, which will detect $\sim500\,000$ galaxies over 75\% of the sky for $\delta<+30^{\circ}$. In the lead up to full telescope operations and the full survey commencing, a number of smaller scale early science, pre-pilot and pilot survey data sets have been observed for verification and validation of the telescope and processing pipeline. WALLABY early science used a subset of 12--16 ASKAP antennas and limited bandwidth to observe four 30-square-degree fields \citep[for details see][]{Reynolds2019, Lee-Waddell2019, Elagali2019, Kleiner2019, For2019}. Pre-pilot observations of a single 30-square-degree field were the first WALLABY observations to use the full 36 antenna array of ASKAP and with its full bandwidth (for details see For et al. in prep.; \citeauthor{Murugeshan2021} \citeyear{Murugeshan2021}). Currently underway is the WALLABY pilot survey, which is observing three 60-square-degree fields in the directions of the Hydra I cluster, Norma cluster and NGC\,4636 group using the full 36 antenna array and full 288\,MHz bandwidth. The purpose of the pilot survey is to provide verification of the full ASKAP system (observing and data processing) and WALLABY observing strategy, thereby demonstrating that ASKAP and its purpose-made software are capable and ready for the full WALLABY survey.

The first pilot survey observations processed were the 60-square-degree field of view covering the Hydra I cluster, which makes Hydra I the first cluster observed as part of the WALLABY survey. These observations provide a rich sample of H\,\textsc{i} detected galaxies in and around Hydra I to investigate the processes affecting galaxies in clusters. 

\subsection{The Hydra I Cluster}
\label{s-sec:hydra_cluster}

The Hydra I cluster \citep[otherwise known as `Abell\,1060' -- see][]{Abell1958} is centred on $\alpha,\delta=$ 10:36:41.8, $-27$:31:28 (J2000), c$z=3\,777$\,km\,s$^{-1}$ (heliocentric reference frame) and appears isolated in redshift \citep[i.e.\ relatively little contamination by foreground/background galaxies,][]{Richter1982}. The CMB (cosmic microwave backgroud) reference frame velocity of the Hydra I cluster is c$z=4\,121$\,km\,s$^{-1}$, which corresponds to a luminosity distance of $D_{\mathrm{lum}}=61$\,Mpc for $H_0=67.7$\,km\,s$^{-1}$\,Mpc$^{-1}$. This agrees with the redshift-independent distance measurement of 59\,Mpc using the Fundamental Plane \citep{Jorgensen1996}. Throughout this work we assume a distance to the Hydra I cluster of 61\,Mpc. It has a catalogued membership of 581 optically detected galaxies \citep{Richter1989}. We adopt the $R_{200}$ size (i.e.\ internal to this radius the mean density is 200 times greater than the critical density of the universe, $\rho_{\mathrm{crit}}$) from \cite{Reiprich2002} as the cluster virial radius of $r_{\mathrm{vir}}\sim1.44$\,Mpc ($\sim1.35^{\circ}$ projected on the sky at 61\,Mpc).

In this work, we provide a first look at the H\,\textsc{i} detections within the virial radius of Hydra I and carry out a case study investigating the possible origin of one cluster member with an asymmetric and disturbed H\,\textsc{i} morphology. We take advantage of the high spatial resolution of WALLABY H\,\textsc{i} data to model the interplay between the gravitational restoring force and ram pressure over the H\,\textsc{i} disc to determine whether ram pressure can explain its H\,\textsc{i} structure. We present the data we use for this work in Section~\ref{sec:data}. In Section~\ref{s-sec:cluster_membership}, we present the first look at Hydra I H\,\textsc{i} detections. Section~\ref{sec:analysis} describes the analysis of our case study galaxy, ESO\,501$-$G075. We discuss environmental mechanisms that could be responsible for the observed H\,\textsc{i} morphology in Section~\ref{sec:discussion} and summarise our conclusions in Section~\ref{sec:conclusion}. Throughout, we adopt optical velocities (c$z$) in the heliocentric reference frame and assume a flat $\Lambda$CDM cosmology with $H_0=67.7$\,km\,s$^{-1}$\,Mpc$^{-1}$ \citep{Planck2016}.

\section{Data}
\label{sec:data}

The WALLABY pilot survey observations of the Hydra I cluster were taken over four nights between 25--26 October and 20--24 November 2019. The raw ASKAP visibility data were flagged, calibrated and imaged using the ASKAP data processing software, ASKAP\textsc{soft} \citep{Whiting2020}, using the standard method \citep[see][for further details]{Reynolds2019, Lee-Waddell2019, Elagali2019, Kleiner2019, For2019}. The output spectral line cubes have a clean 30\,arcsec synthesised beam (emission that was not cleaned because it is below the minor cycle deconvolution threshold of 3.5\,mJy will have a slightly smaller beam size that varies with frequency, see \citeauthor{Reynolds2019} \citeyear{Reynolds2019} and \citeauthor{Kleiner2019} \citeyear{Kleiner2019} for further details on cleaning). Once processed, the 36 beam mosaicked spectral line cubes (denoted footprints A and B with a pointing offset of 0.64 degrees between footprints) were made publicly available from the CSIRO ASKAP Science Data Archive, CASDA\footnote{\url{http://data.csiro.au}} \citep{Chapman2015,Huynh2020}. The footprint A and B cubes were then mosaicked to produce the final full sensitivity cube for H\,\textsc{i} source finding. 

The final spectral line image cube reaches a root-mean-square (RMS) noise of $\sim1.6$\,mJy per beam per 4\,km\,s$^{-1}$ channel. We note that there is no spectral smoothing applied to the data\footnote{Hanning smoothing is not required because ASKAP’s fine filterbank is implemented as a polyphase filterbank, so adjacent channels have very little overlap and can be treated as independent \citep{Hotan2021}.} and the raw channel resolution of 18.5\,kHz corresponds to the final spectral resolution of 4\,km\,s$^{-1}$. This marks the first time WALLABY observations have reached the target WALLABY sensitivity (compared to $\sim2.0$--2.3\,mJy\,beam$^{-1}$ in Early Science, e.g.\ \citeauthor{Reynolds2019} \citeyear{Reynolds2019}, \citeauthor{Lee-Waddell2019} \citeyear{Lee-Waddell2019}). We note that the sensitivity varies from $\sim1.6$--2.2\,mJy across most of the field of view and increases to $\sim3.7$\,mJy near the edges of the field. Source finding was carried out using the Source Finding Application 2 (SoFiA 2, \citeauthor{Serra2015a} \citeyear{Serra2015a}; \citeauthor{Westmeier2021} \citeyear{Westmeier2021}) on a sub-region spanning $\sim3.0^{\circ}\times5.5^{\circ}$, centred on the Hydra I cluster ($\alpha,\delta=$ 10:36:41.8, $-27$:31:28) and covering a velocity range of 500--15\,000\,km\,s$^{-1}$. See \cite{Westmeier2021} for details of the source finding parameters used for running SoFiA 2 on the spectral line cube. The resulting source catalogue, containing 148 H\,\textsc{i} detections, will be included in a future public WALLABY data release containing all WALLABY pilot survey phase 1 observations. For our analysis we use the SoFiA 2 output source data products (e.g.\ source cubelets, moment maps, spectra) and source parameters (e.g.\ position, systemic velocity, integrated flux, position angle).

\section{Cluster Membership}
\label{s-sec:cluster_membership}

\begin{figure}
	\includegraphics[width=\columnwidth]{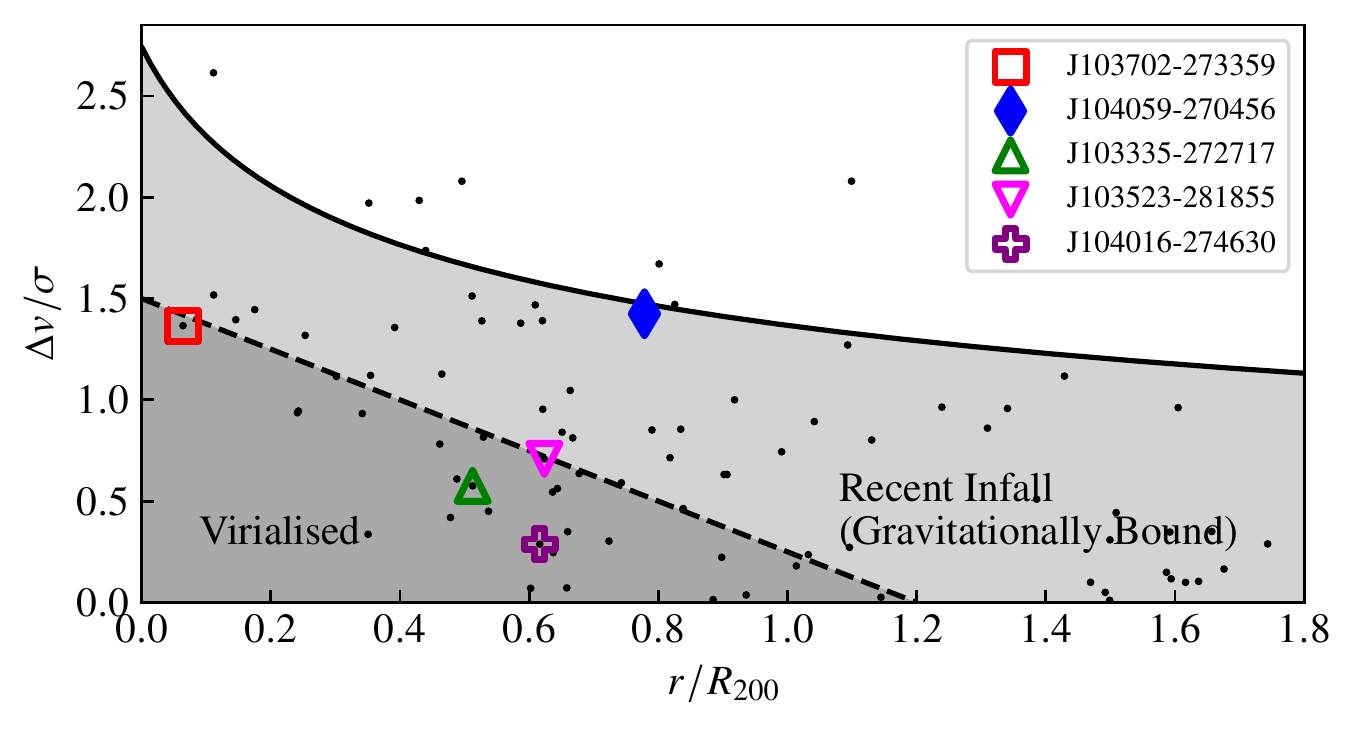}
    \caption{Phase-space diagram for the Hydra I cluster with black points indicating the location of WALLABY H\,\textsc{i} detections. We note that some H\,\textsc{i} detections lie outside the plotted region. The solid black curve indicates the escape velocity for the Hydra I cluster. Galaxies bound to the cluster statistically lie below this curve in the light grey shaded region. Virialised galaxies statistically lie in the dark grey shaded region below the black dashed line. ESO\,501$-$G075 (filled blue diamond) lies slightly below the escape velocity curve, indicating that it is likely on its initial infall into the cluster. The unfilled symbols indicate the location of four other cluster galaxies spatially resolved in H\,\textsc{i} with WALLABY by $>5$ beams (Figure~\ref{fig:5beam_galaxies}).}
    \label{fig:phase_space}
\end{figure}

Not all of the H\,\textsc{i} detections are members of Hydra I, as source finding was carried out over a $\sim15\,000$\,km\,s$^{-1}$ velocity window. We define cluster membership within the virial radius based on the projected cluster phase-space diagram (Figure~\ref{fig:phase_space}), which shows the line of sight velocity difference between galaxies and the cluster normalised by the cluster velocity dispersion, $\Delta v/\sigma$, ($\sigma=676$\,km\,s$^{-1}$, \citeauthor{Richter1982} \citeyear{Richter1982}) versus the galaxies' projected distance from the cluster centre normalised by $R_{200}$, $r/R_{200}$. A galaxy is identified as a member of Hydra I if it lies below the cluster escape velocity curve (i.e.\ it is likely gravitationally bound to the cluster). We derive the escape velocity curve following equations~1--4 from \cite{Rhee2017}, assuming $R_{200}=1.44$\,Mpc, where the corresponding mass enclosed within $R_{200}$ is $M_{200}=3.13\times10^{14}\,\mathrm{M}_{\sun}$ \citep{Reiprich2002}. The projected phase-space diagram can be used to identify galaxies likely to be cluster members as these galaxies will statistically lie within the light and dark grey shaded regions \citep[e.g.][]{Jaffe2015,Yoon2017,Rhee2017}. We note that defining galaxies as cluster members using the phase-space diagram assumes that Hydra I is gravitationally bound and virialised. 

\begin{figure*}
	\includegraphics[width=16cm]{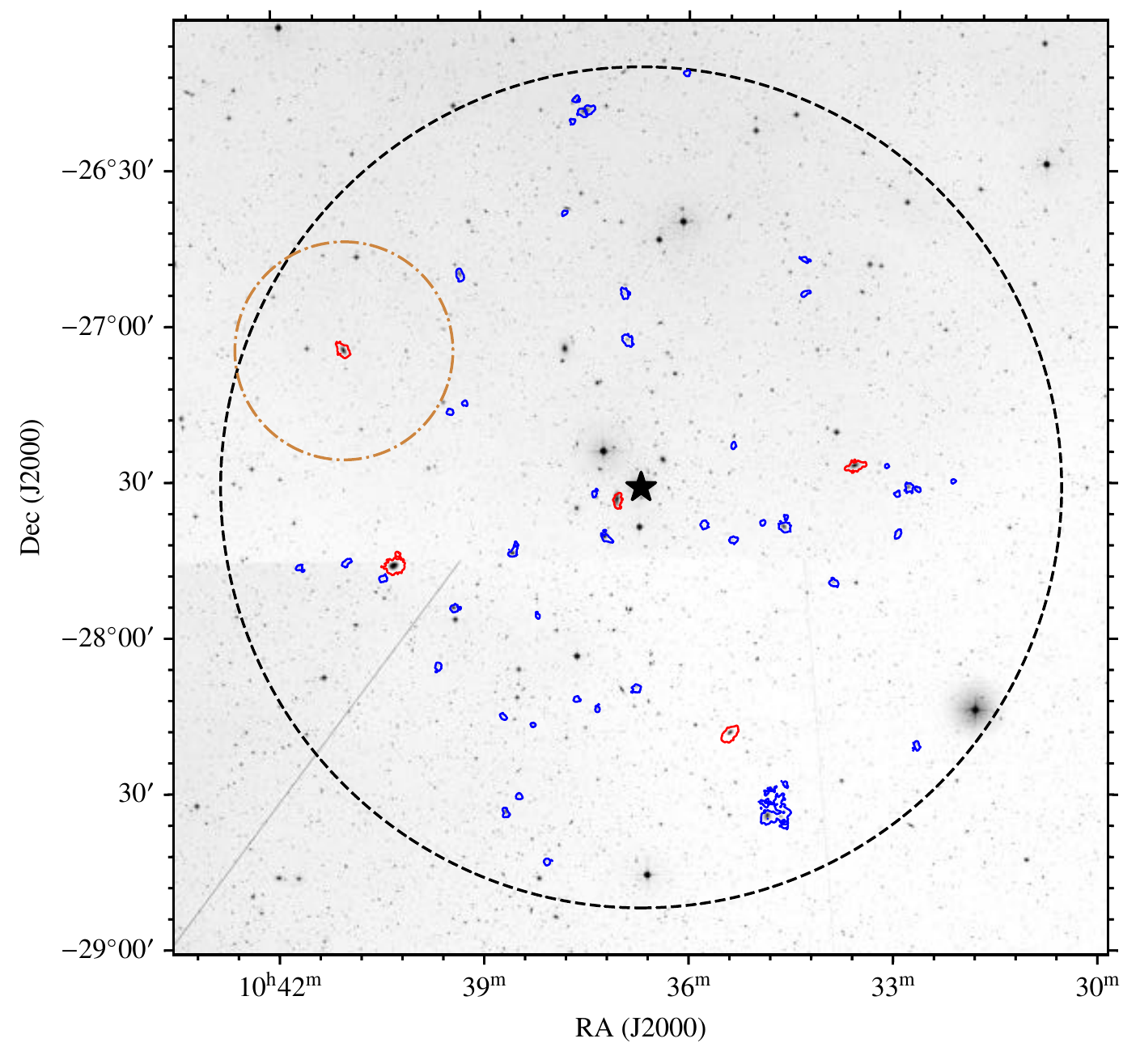}
    \caption{Digitized Sky Survey (DSS) grey-scale image of the Hydra I cluster with ASKAP detections shown by the blue/red contours, indicating an H\,\textsc{i} column density of $5\times10^{19}$\,cm$^{-2}$. ESO\,501$-$G075 is highlighted with the red H\,\textsc{i} column density contour, with the orange dot-dashed circle indicating a projected distance of $0.35^{\circ}$ ($\sim0.37$\,Mpc at a distance of 61\,Mpc) around it. The black star and dashed circle indicate the centre and virial radius ($\sim1.35^{\circ}$, $\sim1.44$\,Mpc) of the Hydra I cluster. We note that the diagonal grey line in the lower left corner is an artefact in the DSS image. We also note the large H\,\textsc{i} contour to the lower right is an interacting galaxy group.}
    \label{fig:sky_map}
\end{figure*}

We find that 51 galaxies are detected in H\,\textsc{i} emission within the virial radius and are members of Hydra I. In Figure~\ref{fig:sky_map}, we show the projected sky distribution of these 51 H\,\textsc{i} detected galaxies. We note that there is not a one-to-one correspondence between galaxies and H\,\textsc{i} contours. In a few instances interacting galaxies are contained within a single H\,\textsc{i} envelope or multiple H\,\textsc{i} detections are part of a single galaxy (commonly referred to as shredding in optical source finding).

\subsection{Resolved Subsample}
\label{s-sec:subsample}

\begin{figure*}
	\includegraphics[width=17.5cm]{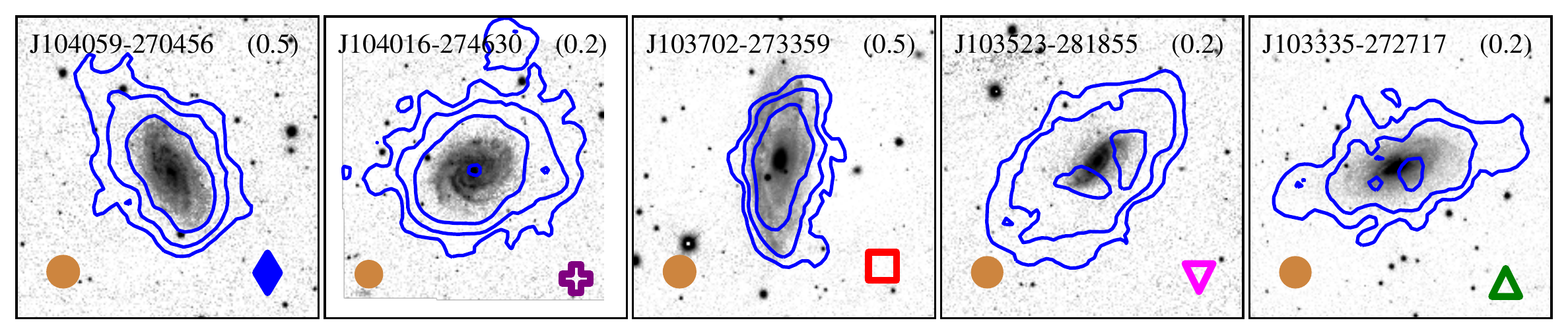}
    \caption{PanSTARRS $g$-band images with overlaid integrated H\,\textsc{i} intensity contours (blue) for Hydra I cluster members resolved by $>5$ beams ($>150''$) along the major axis. The contour levels show H\,\textsc{i} column densities of 5, 20 and $50\times10^{19}$\,cm$^{-2}$. The WALLABY synthesised beam is indicated in the lower left corner by the filled orange circle. The symbols in the lower right corner correspond to the symbols in Figure~\ref{fig:phase_space}. Galaxies are ordered by decreasing RA position on the sky. The number in the top righ corner of each panel is the calculated H\,\textsc{i} morphological asymmetry.}
    \label{fig:5beam_galaxies}
\end{figure*}

Of the 51 H\,\textsc{i} detections classified as cluster members, we identify five galaxies (Figure~\ref{fig:5beam_galaxies} and Table~\ref{table:all_galaxies}) that are spatially resolved by $>5$ beams along the major axis (i.e.\ sufficient for kinematic and mass model analysis, e.g.\ \citeauthor{DiTeodoro2015} \citeyear{DiTeodoro2015}; \citeauthor{Lewis2019} \citeyear{Lewis2019}). We only consider H\,\textsc{i} detections with a single galaxy contained within the H\,\textsc{i} envelope and that have not been shredded by the source finder. We note that the extended H\,\textsc{i} source located near 10:35:00, $-28$:30:00 is an interacting galaxy group and we exclude it from our sample. We indicate these galaxies in Figure~\ref{fig:sky_map} with red contours.

\begin{table*}
	\centering
	\caption{Resolved galaxy subsample properties. Columns are the ESO catalogue name, morphological type, RA ($\alpha$), Dec ($\delta$), heliocentric velocity (c$z$), the line of sight velocity difference relative to the cluster ($\Delta$c$z_{\mathrm{cl}}$), the luminosity distance set to that of the cluster ($D_{\mathrm{L}}$), the morphological asymmetry ($A_{\mathrm{map}}$) and the number of neighbouring galaxies within a projected distance of $<0.35^{\circ}$ and velocity difference of $<500$ (1\,000)\,km\,s$^{-1}$ ($N_{\mathrm{neighbour}}$). We highlight J104059$-$270456 in bold as it is the subject of this work.}
	\label{table:all_galaxies}
	\begin{tabular}{lccccccccr}
		\hline
		WALLABY & ESO & Type & $\alpha$ & $\delta$ & c$z$ & $\Delta$c$z_{\mathrm{cl}}$  & $D_{\mathrm{L}}$ & $A_{\mathrm{map}}$ & $N_{\mathrm{neighbour}}$ \\
		& & & \multicolumn{2}{c}{[J2000]} & [km\,s$^{-1}$] & [km\,s$^{-1}$] & [Mpc] & & \\
		\hline
		\textbf{J104059$-$270456} & 501$-$G075 & SA(s)c & 10:40:59 & $-27$:04:56 & 4740 & $\sim960$ & 61 & 0.5 & 0 (0) \\
		J103702$-$273359 & 501$-$G043 & SA(s)b & 10:37:02 & $-27$:33:59 & 2854 & $\sim930$ & 61 & 0.5 & 11 (20) \\
		J103335$-$272717 & 501$-$G015 & SB(s)a & 10:33:35 & $-27$:27:17 & 3388 & $\sim390$ & 61 & 0.2 & 10 (12) \\
		J103523$-$281855 & 437$-$G004 & SAB(r)bc & 10:35:23 & $-28$:18:55 & 3298 & $\sim480$ & 61 & 0.2 & 13 (15) \\
		J104016$-$274630 & 437$-$G036 & SAB(rs)c & 10:40:16 & $-27$:46:30 & 3972 & $\sim190$ & 61 & 0.2 & 6 (8) \\
		\hline
	\end{tabular}
\end{table*}

Visually, these galaxies display a diversity of H\,\textsc{i} morphologies. We quantify the level of disturbance in the H\,\textsc{i} morphologies by the asymmetry in the moment 0 map. The map asymmetry, $A_{\mathrm{map}}$, \citep[first defined for measuring optical asymmetries by][]{Abraham1996} is defined as the sum of residuals between the moment 0 map and itself rotated by $180^{\circ}$ about the centre of mass of the galaxy. We minimise $A_{\mathrm{map}}$ \citep[e.g.][]{Conselice2000} by varying the rotation centre by $\pm5$ pixels (i.e.\ $\pm1$ beam) in RA and Dec of the centre of mass of the moment 0 map. We have ordered the galaxies in Figure~\ref{fig:5beam_galaxies} from most to least asymmetric. We find a clear split in this sample, with the first two galaxies both having large asymmetries ($A_{\mathrm{map}}\sim0.5$), whereas the last three galaxies are fairly symmetric ($A_{\mathrm{map}}\sim0.2$). We note that, as these galaxies have similar angular and physical resolutions and local noise levels in the spectral cube, many of the systematic effects discussed by \cite{Giese2016} do not affect the relative comparison of $A_{\mathrm{map}}$ values. However, these asymmetries are not directly comparable with other surveys.

The high asymmetry of J103702$-$273359 (first panel of Figure~\ref{fig:5beam_galaxies}) is likely to be caused by ram pressure stripping, based on the truncation of the H\,\textsc{i} disc within the optical radius on the Northern side and compression (puffing out) of the H\,\textsc{i} on the Eastern (Western) side. This is further supported by J103702$-$273359 being located near the cluster centre (red contour near the star, which indicates the centre of the Hydra I cluster in Figure~\ref{fig:sky_map}; $r/R_{200}<0.1$, red square in Figure~\ref{fig:phase_space}). J104059$-$270456 (second panel of Figure~\ref{fig:5beam_galaxies}) displays an H\,\textsc{i} tail on its North-East side (upper left). Although J103335$-$272717 and J104016$-$274630 (third and fifth panels of Figure~\ref{fig:5beam_galaxies}, respectively) have relatively low asymmetries of $A_{\mathrm{map}}\sim0.2$, we note low level signs of disturbed H\,\textsc{i} at their outer edges that can be explained by close neighbours (see Table~\ref{table:all_galaxies}). We quantify the local density around each galaxy by calculating the number of nearby neighbours with projected angular separations of $<0.35^{\circ}$ and velocity differences of $<500$ and $<1\,000$\,km\,s$^{-1}$ (Table~\ref{table:all_galaxies}) from the 6dF Galaxy Survey catalogue \citep{Jones2009}. ESO\,501$-$G075 is the only one of the five galaxies with no close neighbours. This suggests that its morphology in unlikely to be caused by tidal interactions with close companions.

\subsubsection{Phase-Space}
\label{ss-sec:phasespace}

We show the location of these five galaxies on the cluster phase-space diagram in Figure~\ref{fig:phase_space}. Two of these galaxies, J103335$-$272717 and J104016$-$274630 (green triangle and purple cross, respectively), lie within the dark grey shaded region within which galaxies are statistically virialised. However, neither galaxy has a truncated H\,\textsc{i} disc, so it is more likely that these galaxies are falling into the cluster with low line of sight velocities relative to that of the cluster. J103523$-$281855 and J103702$-$273359 (pink triangle and red square, respectively) lie near the border between the regions that statistically contain galaxies that are virialised or recent infalls (below and above dashed black line, respectively). J103523$-$281855 (pink triangle) has similar properties (fairly symmetric with a more extended H\,\textsc{i} disc relative to its optical disc) to J103335$-$272717 and J104016$-$274630. 

J104059$-$270456 (filled blue diamond) stands out as residing in a distinctly different part of phase-space with the largest projected distance from the cluster centre ($r/R_{200}\sim0.8$) and has a normalised velocity difference near the cluster escape velocity (solid black curve). Of these five galaxies, J104059$-$270456 appears to be the at the earliest stage of infalling into the cluster and yet has an asymmetry similar to J103702$-$273359, which is at a cluster centric distance of $r/R_{200}<0.1$ and has likely spent more time in the cluster. J104059$-$270456 is an excellent candidate for probing environmental processing acting on a galaxy on its initial infall into the cluster environment and we use this galaxy as a case study of what WALLABY will be able to achieve for statistical samples of galaxies. In the rest of this paper we investigate whether ram pressure could be responsible for the observed H\,\textsc{i} morphology.

\section{Analysis}
\label{sec:analysis}

\subsection{ESO501-G075}
\label{s-sec:eso501-g075}

We find the WALLABY source J104059$-$270456 to be associated with the optical counterpart ESO\,501$-$G075 (angular separation $\sim6.5$\,arcsec). ESO\,501$-$G075 is a spiral galaxy with morphological type SA(s)c \citep{devaucouleurs1991} located at $\alpha,\delta=$ 10:40:59, $-27$:04:59 (J2000). ESO\,501$-$G075 lies to the North-East, NE, of the cluster centre at a projected distance of $\sim1.05^{\circ}$ ($\sim1.12$\,Mpc), which is $\sim0.3^{\circ}$ ($\sim0.32$\,Mpc) inside the virial radius and has no nearby neighbours with optical or H\,\textsc{i} detections within a projected distance of $\sim0.35^{\circ}$ ($\sim0.37$\,Mpc, Figure~\ref{fig:sky_map}) and velocity difference of 1\,000\,km\,s$^{-1}$. We quantify the asymmetry in the stellar distribution of ESO\,501$-$G075 by calculating the CAS morphometric parameters \citep[concentration, asymmetry, smoothness,][respectively]{Bershady2000,Conselice2000,Conselice2003} using the \textsc{python} package \textsc{statmorph} \citep{Rodriguez2019}. ESO\,501$-$G075 appears undisturbed in the optical $g$-band image ($C=2.41$, $A=0.03$, $S=0.02$).

The systemic barycentric velocity of ESO\,501$-$G075 is $\mathrm{c}z=4\,740$\,km\,s$^{-1}$, which gives it a velocity difference relative to the centre of the Hydra I cluster of $\Delta\mathrm{c}z_{\mathrm{cl}}\sim960$\,km\,s$^{-1}$. ESO\,501$-$G075 has one of the highest stellar and H\,\textsc{i} masses of galaxies in the Hydra I cluster detected in H\,\textsc{i} with a stellar mass of $\log(M_*/[\mathrm{M}_{\sun}])=10.36\pm0.13$ (derived from VHS\footnote{VISTA Hemisphere Survey \citep{McMahon2013}.} $J$- and $K$-band photometry, see \citeauthor{Reynolds2019} \citeyear{Reynolds2019} for details) and an H\,\textsc{i} mass of $\log(M_\mathrm{HI}/[\mathrm{M}_{\sun}])=9.75\pm0.02$. Table~\ref{table:galaxy_properties} summarises optical and H\,\textsc{i} derived galaxy properties of ESO\,501$-$G075.

\begin{table}
	\centering
	\caption{ESO\,501$-$G075 properties.}
	\label{table:galaxy_properties}
	\begin{tabular}{lr}
		\hline
		Parameter & Value \\
		\hline
		$S_{\mathrm{int}}$ [Jy\,km\,s$^{-1}$] & $6.54\pm0.26$ \\ 
	    $w_{20}$ [km\,s$^{-1}$] & 406 \\
		$\log(M_*/[\mathrm{M}_{\sun}])$ & $10.36\pm0.13$ \\
		$\log(M_{\mathrm{HI}}/[\mathrm{M}_{\sun}])$ & $9.75\pm0.02$ \\
		$\log(M_{\mathrm{Dyn}}/[\mathrm{M}_{\sun}])$ & $11.3\pm0.1$ \\
		$\log(M_{\mathrm{HI}}/M_*)$ & $-0.6$ \\
		$\mathrm{DEF}_{\mathrm{HI}}$ & $0.14$ \\
		$A_{\mathrm{flux}}$ & 1.03 \\
		$d_{25}$ [kpc] & 35.1 \\
		\hline
	\end{tabular}
\end{table}

\subsection{H\textsc{i} Content}
\label{s-sec:hi_prop}

\begin{figure}
	\includegraphics[width=\columnwidth]{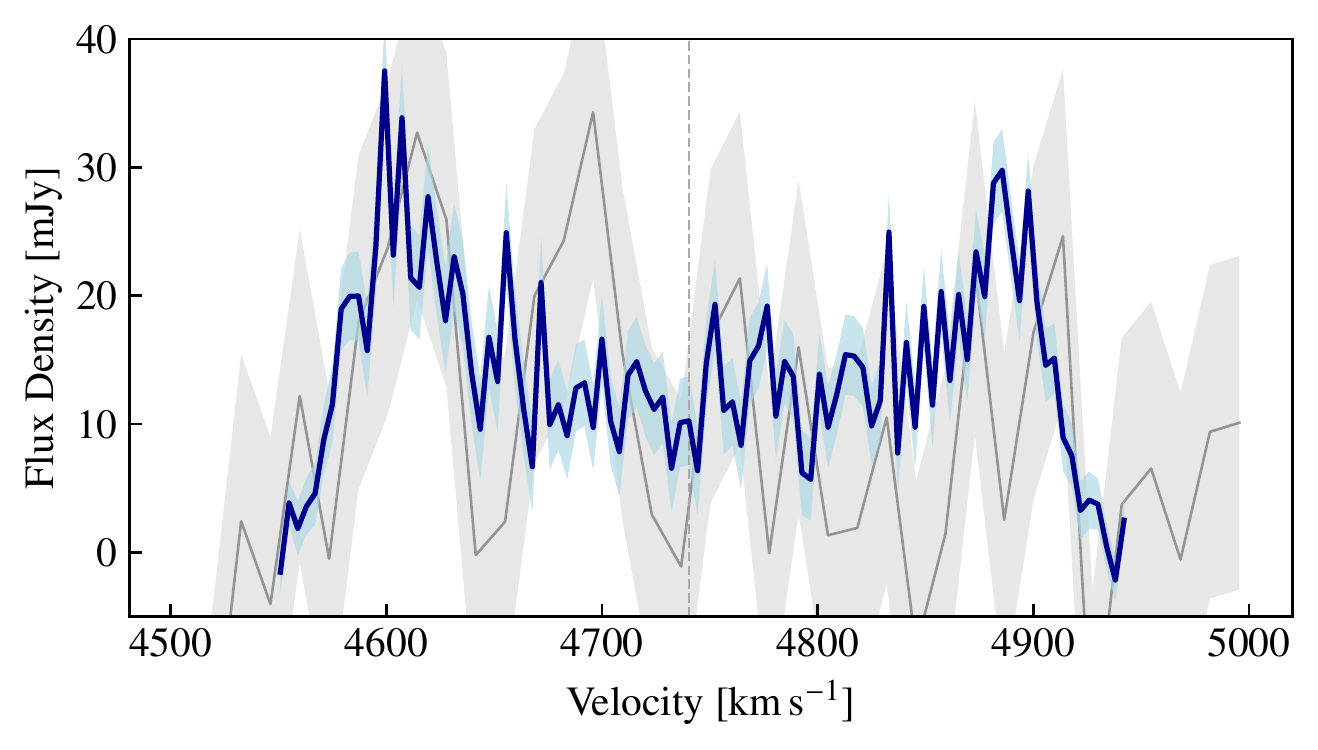}
    \caption{The integrated H\,\textsc{i} spectra for ESO\,501$-$G075 from ASKAP (blue) and extracted from HIPASS (grey). The light shaded regions indicate the integrated noise in each channel.}
    \label{fig:spectrum}
\end{figure}

We show the integrated H\,\textsc{i} spectrum from ASKAP (blue) for ESO\,501$-$G075 in Figure~\ref{fig:spectrum}. Although ESO\,501$-$G075 is not listed in the HIPASS source catalogue \citep[HICAT,][]{Meyer2004}, we also extract a spectrum from the HIPASS spectral cube covering the same area of sky, which we show in grey. The light blue and grey shaded regions indicate the integrated noise in each ASKAP channel and the per channel noise for HIPASS ($\sigma_{\mathrm{HIPASS}}=13$\,mJy\,beam$^{-1}$), respectively, which illustrates why ESO\,501$-$G075 was not detected in HIPASS. The integrated flux from the HIPASS spectrum is 4.0\,Jy\,km\,s$^{-1}$ with an integrated signal to noise ratio (SNR) of $\sim4.8$ (i.e.\ below the HICAT SNR limit of 5).

\begin{figure}
	\includegraphics[width=\columnwidth]{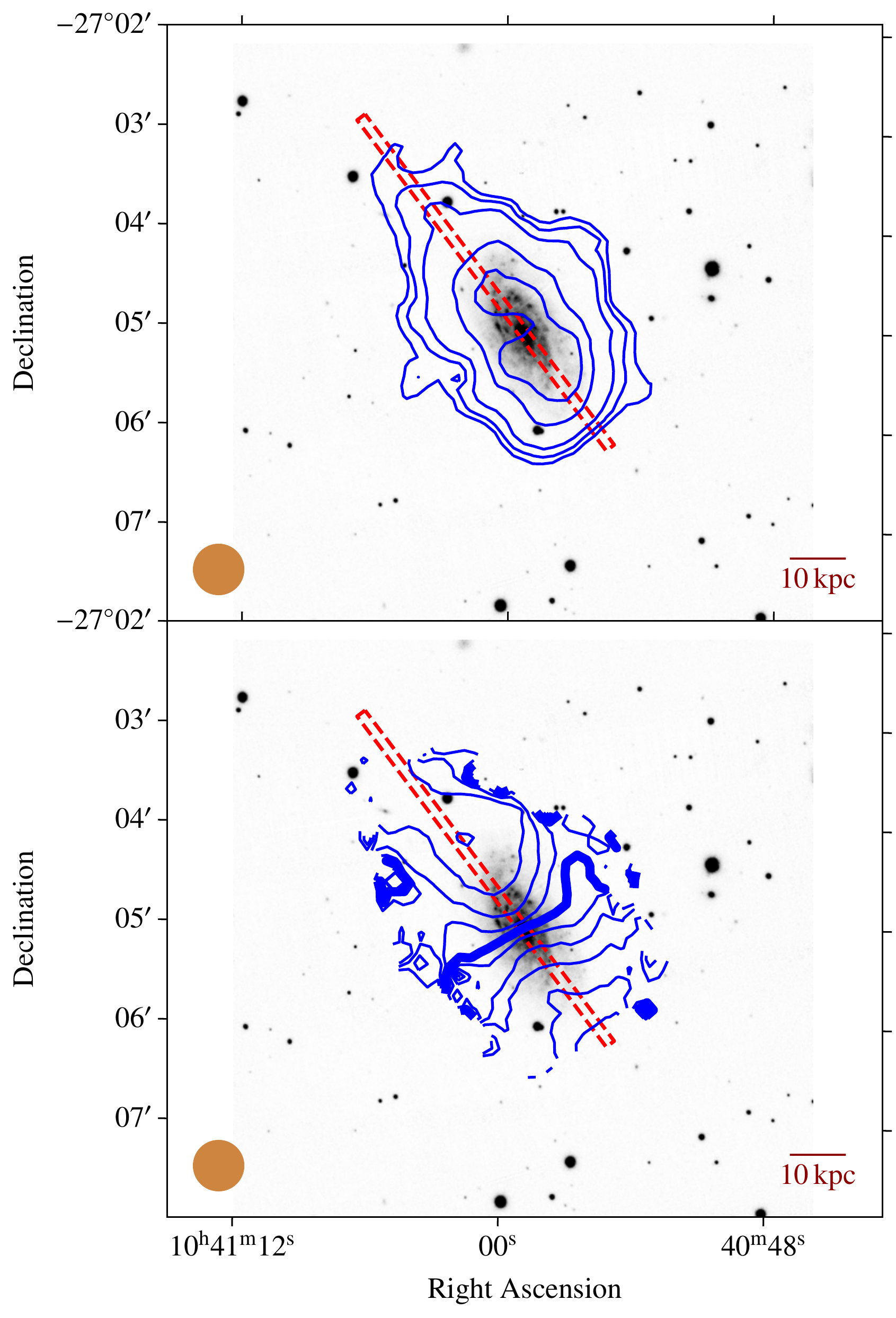}
    \caption{Grey-scale Pan-STARRS $g$-band image of ESO\,501$-$G075 with overlaid ASKAP H\,\textsc{i} column density contours (upper panel): 5, 10, 20, 50, $70\times10^{19}$\,cm$^{-2}$ and velocity field contours (lower panel): steps of 40\,km\,s$^{-1}$ from the systemic velocity (c$z=4740$\,km\,s$^{-1}$, thick contour). The filled orange circle in the lower left corner shows the ASKAP synthesised beam size. The red dashed rectangle indicates the box used to extract the position-velocity diagram and column density profile shown in Figures~\ref{fig:pv_diagram} and \ref{fig:colden_cut}, respectively. The line in the lower right corner indicates a physical scale of 10\,kpc for a distance to ESO\,501$-$G075 of 61\,Mpc.}
    \label{fig:moment0_map}
\end{figure}

In the upper panel of Figure~\ref{fig:moment0_map}, we show a grey-scale Pan-STARRS \citep{Chambers2016,Flewelling2016} $g$-band image of ESO\,501$-$G075 with overlaid contours (blue) showing H\,\textsc{i} column densities from ASKAP. We overlay velocity field contours in the lower panel of Figure~\ref{fig:moment0_map}. Unlike the optical disc, the H\,\textsc{i} gas disc (H\,\textsc{i} integrated intensity, moment 0, map) shows signs of a disturbed H\,\textsc{i} morphology through the compression (stretching) of the H\,\textsc{i} contours on the South-West, SW, (North-East, NE) side of the galaxy. This suggests that ESO\,501$-$G075 may be experiencing one or more environmental processes that are disturbing the H\,\textsc{i}, but are not affecting the stellar component. We next look for additional evidence that ESO\,501$-$G075 is disturbed by the environment using integrated H\,\textsc{i} properties: H\,\textsc{i} deficiency (Section~\ref{ss-sec:hi_def}) and H\,\textsc{i} spectral asymmetry (Section~\ref{ss-sec:hi_asym}).

\subsubsection{H\textsc{i} Deficiency}
\label{ss-sec:hi_def}

Galaxies in dense, cluster environments are more frequently found to be H\,\textsc{i} deficient compared to counterparts in the field \citep[e.g.][]{Solanes2001, Chung2009, Boselli2009, Brown2017} with similar optical/stellar properties. Such H\,\textsc{i} deficiencies are likely caused by the external mechanisms observed to be acting in these denser environments \citep[e.g.\ ram pressure stripping and tidal interactions,][]{BravoAlfaro2000, Kenney2004, Chung2007, Yoon2017}. Semi-analytic and hydrodynamical simulations also show that ram pressure stripping leads to galaxies that are more H\,\textsc{i} poor \citep[e.g.][]{Stevens2017, Stevens2019}. Stripping a substantial amount of H\,\textsc{i} from a galaxy generally requires the galaxy to have traversed the cluster or been within the cluster for a significant length of time (e.g.\ $\sim10^7$--$10^8$\,yr for ram pressure, \citeauthor{Abadi1999} \citeyear{Abadi1999}; \citeauthor{Vollmer2001} \citeyear{Vollmer2001}, and $\sim10^9$\,yr for viscous stripping, \citeauthor{Nulsen1982} \citeyear{Nulsen1982}; \citeauthor{Quilis2000} \citeyear{Quilis2000}). Hence, estimating the H\,\textsc{i} deficiency of ESO\,501$-$G075 can provide an indication of whether ESO\,501$-$G075 has interacted with the environment in the past.

We derive the expected H\,\textsc{i} mass of ESO\,501$-$G075 from the optical size and galaxy morphology using the relation \citep[e.g.][]{Haynes1984, Cortese2011} given by,
\begin{equation}
	\log(M_{\mathrm{HI,exp}}/[\mathrm{M}_{\sun}])=a + 2b \times \log\left(\frac{hd_{25}}{[\mathrm{kpc}]}\right) - 2\log(h),
	\label{equ:mhi_exp}
\end{equation}
where $a=7.16$ and $b=0.87$ are constants corresponding to the morphological type Sc from table 3 of \cite{Boselli2009}, $d_{25}=35.1$\,kpc is the $B$-band optical 25$^{\mathrm{th}}$\,mag\,arcsec$^{-2}$ diameter (calculated from NED\footnote{NASA/IPAC Extragalactic Database.}) and $h=H_0/(100$\,km\,s$^{-1})$. Using the standard definition, the H\,\textsc{i} deficiency, $\mathrm{DEF}_{\mathrm{HI}}$, is,
\begin{equation}
	\mathrm{DEF}_{\mathrm{HI}} = \log(M_{\mathrm{HI,exp}}/[\mathrm{M}_{\sun}]) - \log(M_{\mathrm{HI,obs}}/[\mathrm{M}_{\sun}]).
	\label{equ:def_hi}
\end{equation}
Galaxies are generally considered to be H\,\textsc{i} deficient if $\mathrm{DEF}_{\mathrm{HI}}>0.3$ and H\,\textsc{i} rich if $\mathrm{DEF}_{\mathrm{HI}}<-0.3$ \citep[i.e.\ contain less than half/more than double the expected H\,\textsc{i} content, respectively, e.g.][]{Kilborn2009,Boselli2009}. We find ESO\,501$-$G075 to have a typical amount of H\,\textsc{i}, where $\mathrm{DEF}_{\mathrm{HI}}=0.14$. There are two possible scenarios that can explain this $\mathrm{DEF}_{\mathrm{HI}}$. One is that ESO\,501$-$G075 has not previously experienced a significant external interaction that would have resulted in significant gas stripping. Alternatively, if the galaxy were previously H\,\textsc{i} rich, then ESO\,501$-$G075 could have had gas stripped to become H\,\textsc{i} normal.

\subsubsection{H\textsc{i} Asymmetry}
\label{ss-sec:hi_asym}

Disturbances in a galaxy's H\,\textsc{i} content can be observed in both the spatially resolved H\,\textsc{i} morphology and its integrated H\,\textsc{i} spectrum \citep[e.g.][]{Richter1994}. Studies of the asymmetry in the integrated H\,\textsc{i} spectra of galaxies find that the fraction of asymmetric galaxies increases with increasing environment density \citep[e.g.][]{Richter1994, Haynes1998, Matthews1998, Espada2011, Scott2018, Bok2019, Reynolds2020, Watts2020}. This suggests that external environmental processes (e.g.\ ram pressure stripping and tidal interactions) could be responsible for observed asymmetries. 

The asymmetry in the integrated spectrum is commonly quantified by the flux asymmetry ratio, $A_{\mathrm{flux}}$ \citep[e.g.][]{Richter1994, Haynes1998}. The flux asymmetry ratio is defined as the ratio of the integrated flux in the lower ($I_1$) and upper ($I_2$) halves of the H\,\textsc{i} spectrum split at the galaxy's systemic velocity (e.g.\ for a perfectly symmetric spectrum $A_{\mathrm{flux}}=1$). ESO\,501$-$G075 has a symmetric integrated spectrum with $A_{\mathrm{flux}}=1.03$. The elongated tail on the NE side of ESO\,501$-$G075 contains an H\,\textsc{i} mass of $\sim4.5\times10^8\,\mathrm{M}_{\sun}$ (i.e.\ $\sim6\%$ of the total H\,\textsc{i} mass). The small H\,\textsc{i} mass located in the tail explains why we do not find an asymmetry in the integrated spectrum corresponding to the morphological asymmetry ($A_{\mathrm{map}}=0.5$), as the tail contributes very little to the galaxy's spectrum. 

Based on ESO\,501$-$G075's integrated quantities of H\,\textsc{i} deficiency and spectral asymmetry, the galaxy appears to be unaffected by its projected location with respect to the cluster. This could suggest that ESO\,501$-$G075 lies well beyond the cluster virial radius and is simply seen in projection to be within the virial radius. However, the spatially resolved map indicates that the cluster environment may have begun to affect the gas in ESO\,501$-$G075.

\subsection{Gas Kinematics}
\label{s-sec:kinematics}

We derive the rotation curve of ESO\,501$-$G075 from which we can estimate the dynamical mass. The rotation curve is also a required input for mass modelling to estimate the dark matter distribution \citep[e.g.][]{Chemin2006,Westmeier2011,Reynolds2019,Elagali2019}. We do this by fitting a tilted ring model to the galaxy \citep{Rogstad1974}. Tilted ring modelling assumes that gas particles follow circular orbits and move with constant speed. Under these assumptions, the velocity field can be modelled by a series of circular isovelocity rings with position angles and inclinations that match the galaxy. For this reason, we exclude the extended H\,\textsc{i} tail from our tilted ring model by only fitting out to 82.5\,arcsec as the gas in the tail is likely to be dominated by non-circular motions. Tilted ring modelling requires sufficient spatial resolution (e.g.\ $\gtrsim5$ beams across the major axis). ESO\,501$-$G075 satisfies this requirement with a spatial resolution of $\sim6$ beams across the major axis (Figure~\ref{fig:moment0_map}). 

We use the tilted ring modelling code \textsc{3dbarolo} \citep{DiTeodoro2015} to derive the rotation curve for ESO\,501$-$G075. Unlike some tilted ring modelling codes, which operate on 2-dimensional velocity fields \citep[e.g. the \textsc{gipsy} task \textsc{rotcur},][]{vanderHulst1992,Vogelaar2001}, \textsc{3dbarolo} derives the rotation curve using the full 3-dimensional H\,\textsc{i} spectral line cube. We perform the tilted ring modelling using a cube smoothed to a spectral resolution of 12\,km\,s$^{-1}$ to increase the SNR. We use the \textsc{3dbarolo} mask option \textsc{smooth} with default parameters (i.e.\ spatially smoothing by a factor of 2 and masking pixels after smoothing with $\mathrm{SNR}<3$). We run \textsc{3dbarolo} in two iterations using a set of 6 rings with a width of 15\,arcsec (i.e.\ half the ASKAP synthesised beam, 30\,arcsec). In the first run we leave free the $x,y$ position centre, inclination angle ($i$), position angle ($\theta$), velocity dispersion ($\sigma_{\mathrm{gas}}$) and rotational velocity ($v_{\mathrm{rot}}$). We use the SoFiA source parameters for our initial inputs, except for the inclination, which we set to $i=53^{\circ}$ using the optical $B$-band 25$^{\mathrm{th}}$\,mag\,arcsec$^{-2}$ diameter from NED. We also tested leaving the systemic velocity ($v_{\mathrm{vsys}}$) free, however this produced a poor fit and we chose to fix $v_{\mathrm{vsys}}=4740$\,km\,s$^{-1}$ (the value from SoFiA). For the second run we only leave the velocity dispersion ($\sigma_{\mathrm{gas}}$) and rotational velocity ($v_{\mathrm{rot}}$) as free parameters and fix the $x,y$ position centre, inclination angle and position angle to the mean values from the first run (Table~\ref{table:fixed_params}). We note that we round $i$ and $\theta$ to the nearest degree in agreement with our uncertainty (e.g.\ $\pm2^{\circ}$). Table~\ref{table:fit_params} lists the final parameters from the second iteration of the tilted ring fit. We do not include $\sigma_{\mathrm{gas}}$ in Table~\ref{table:fit_params} as the values are not reliable, but do not significantly affect the final rotation curve. We also note that the uncertainties in the rotation curve are driven by the modelling and are under-estimates of the true model uncertainties.

\begin{table}
	\centering
    \caption{Tilted ring model parameters derived from the first fit and fixed in the second fit. Parameters: ($x,y$) - (RA, Dec), $v_{\mathrm{sys}}$ - systemic velocity, $i$ - inclination, $\theta$ - position angle. The systemic velocity is fixed to the value from the SoFiA source finding.}
	\label{table:fixed_params}
	\begin{tabular}{lcccr}
		\hline
		$x$ & $y$ & $v_{\mathrm{sys}}$ & $i$ & $\theta$ \\
        {h:m:s} & d:m:s & [km\,s$^{-1}$] & [deg] & [deg] \\ \hline
	    10:40:59.2 & $-27$:05:02.4 & 4740 & $57\pm2$ & $214\pm2$  \\
		\hline
	\end{tabular}
\end{table}

\begin{table}
	\centering
    \caption{Tilted ring model fit parameters and errors, and gaseous and stellar mass surface densities and standard deviations for ESO\,501$-$G075. Parameters: $r$~-~radius, $v_{\mathrm{rot}}$ - rotational velocity, $\Sigma_{\mathrm{star}}$ - average stellar mass surface density from VHS $J$- and $K$-bands, $\Sigma_{\mathrm{gas}}$ - gas mass surface density.}
	\label{table:fit_params}
	\begin{tabular}{lcccr}
		\hline
		$r$ & $r$ & $v_{\mathrm{rot}}$ & $\Sigma_{\mathrm{star}}$ & $\Sigma_{\mathrm{gas}}$ \\
        {[arcsec]} & [kpc] & [$\mathrm{km}\,\mathrm{s}^{-1}$] & [$\mathrm{M}_{\sun}\,\mathrm{pc}^{-2}$] & [$\mathrm{M}_{\sun}\,\mathrm{pc}^{-2}$] \\ \hline
		7.5   & 2.2  & $103\pm16$ & $124.4\pm57.9$ & $5.0\pm0.3$ \\
		22.5  & 6.7  & $152\pm10$ & $64.6\pm23.7$  & $5.4\pm0.4$ \\
		37.5  & 11.1 & $174\pm5$  & $20.3\pm16.2$  & $5.0\pm0.4$ \\
		52.5  & 15.5 & $191\pm5$  & $6.8\pm10.1$   & $4.0\pm0.5$ \\
		67.5  & 20.0 & $197\pm5$  & $0.2\pm1.3$    & $2.6\pm0.5$ \\
		82.5  & 24.4 & $196\pm8$  & ---            & $1.3\pm0.5$ \\
		97.5  & 28.8 & ---        & ---            & $0.7\pm0.3$ \\
		112.5 & 33.3 & ---        & ---            & $0.3\pm0.2$ \\
		127.5 & 37.7 & ---        & ---            & $0.2\pm0.1$ \\
		142.5 & 42.1 & ---        & ---            & $0.2\pm0.1$ \\
		157.5 & 46.6 & ---        & ---            & $0.1\pm0.1$ \\
		\hline
	\end{tabular}
\end{table}

\begin{figure}
	\includegraphics[width=\columnwidth]{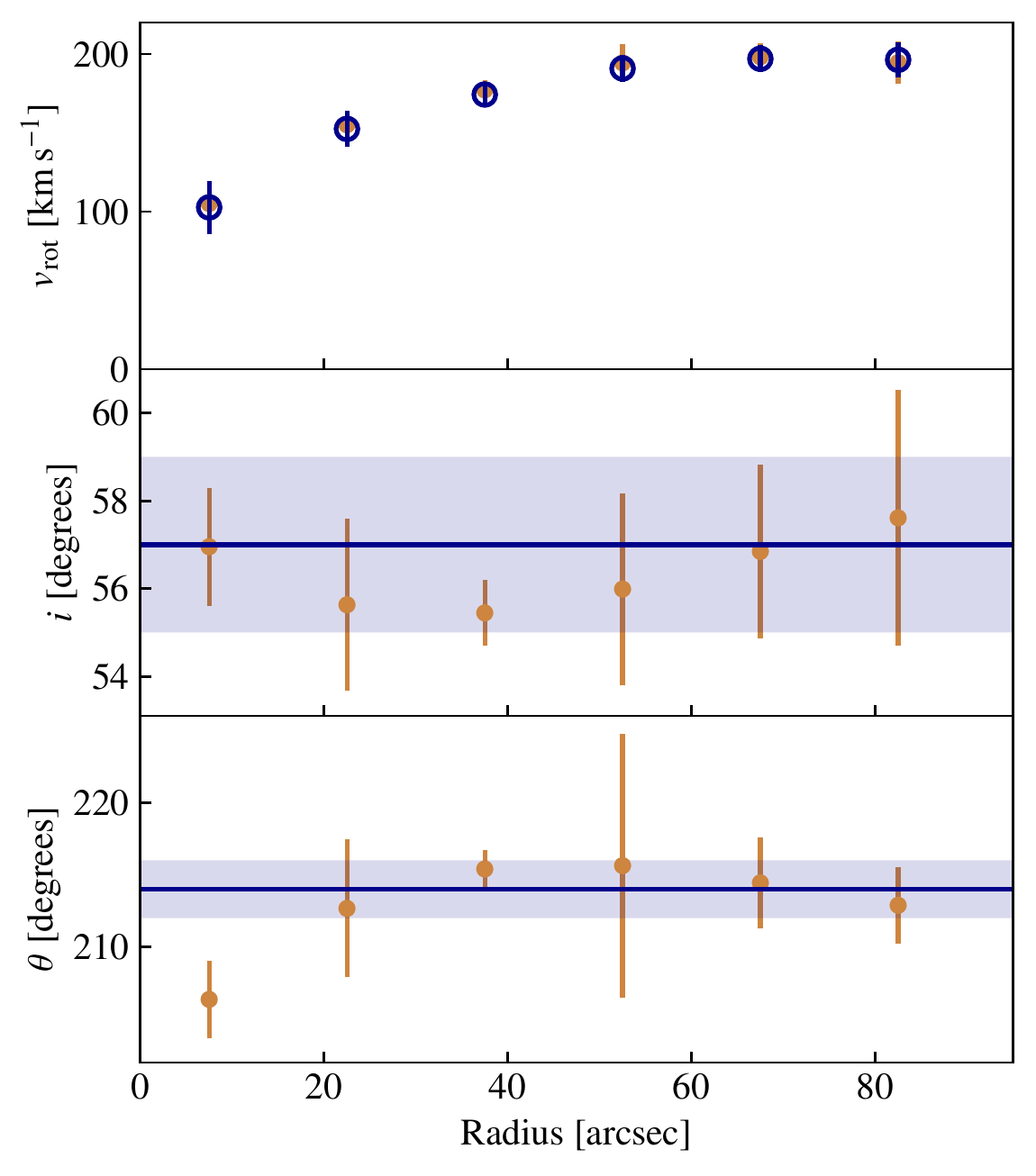}
    \caption{The output rotational velocity ($v_{\mathrm{rot}}$), inclination angle ($i$), and position angle ($\theta$) from the tilted ring modelling (from top to bottom panels). The fitted parameters from the first run of \textsc{3dbarolo} leaving all parameters free are in orange. The final $v_{\mathrm{rot}}$ from the second run is in blue with the fixed values used for $i$ and $\theta$ shown by the horizontal blue lines in the lower two panels with the shaded band indicating the uncertainty. The derived $v_{\mathrm{rot}}$ does not change within the uncertainties between the fits using variable vs fixed $i$ and $\theta$.}
    \label{fig:rotcur_params}
\end{figure}

\begin{figure}
	\includegraphics[width=\columnwidth]{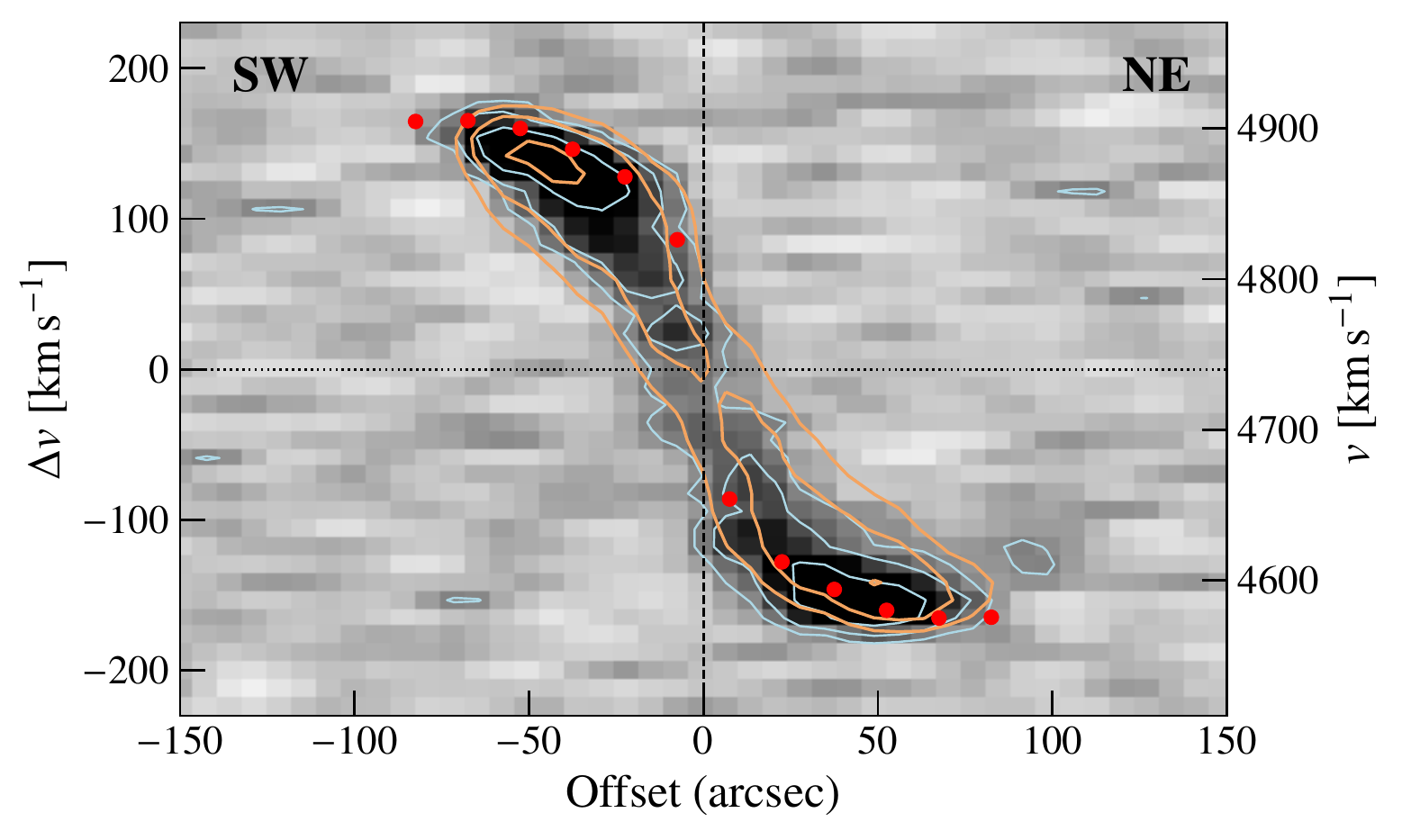}
    \caption{Position-velocity diagram along the major axis of ESO\,501$-$G075 (red rectangle in Figure~\ref{fig:moment0_map}). The blue and orange contours correspond to the data and the tilted ring model, respectively, which show good agreement. The contours correspond to 2, 4 and 8\,mJy per beam per 4\,km\,s$^{-1}$ channel. The filled red circles indicate the derived rotation curve (top panel of Figure~\ref{fig:rotcur_params}). Both the model PV diagram contours and derived rotation curve show good agreement with the data.}
    \label{fig:pv_diagram}
\end{figure}

In Figure~\ref{fig:rotcur_params}, we show the output rotational velocity, inclination angle and position angle (panels from top to bottom, respectively) from the first run of \textsc{3dbarolo} in orange (all parameters free) and the second run in blue (only $v_{\mathrm{rot}}$ and $\sigma_{\mathrm{gas}}$ free). We find that the derived rotation curve is robust with respect to small changes ($\pm2^{\circ}$) in $i$ and $\theta$ as the derived $v_{\mathrm{rot}}$ shows only minor variation between allowing $i$ and $\theta$ to vary with radius and keeping them fixed for all radii. The derived rotation curve appears to be typical of Sc galaxies, which tend to have rotational velocities in the range $v_{\mathrm{rot}}\sim100$--200\,km\,s$^{-1}$ \citep[e.g.][]{Sofue1999}. To assess the goodness of the tilted ring model, we compare the derived rotation curve and \textsc{3dbarolo} model galaxy position-velocity (PV) diagram with the PV diagram extracted along the major axis (red rectangle in Figure~\ref{fig:moment0_map}) of ESO\,501$-$G075 (Figure~\ref{fig:pv_diagram}). We find that the model PV diagram and derived rotation curve are well matched to the PV diagram from the data. We note that, should ESO\,501$-$G075 be undergoing strong ram pressure stripping, then the tilted ring model fit would have large uncertainties due to non-circular motions introduced into the gas disc by the ram pressure. However, we do not believe this to be the case (see Sections~\ref{s-sec:eso501-g075}, \ref{s-sec:hi_prop}, \ref{sec:discussion}). 

\subsection{Mass Modelling}
\label{s-sec:mass_model}

In order to investigate whether ram pressure can effectively strip gas from ESO\,501$-$G075, we require an estimate of the galaxy's gravitational restoring force acting upon the gas. This restoring force is produced by the stars, gas and dark matter comprising the galaxy. Unlike the observable baryonic components (stars and gas), which we can directly measure, we can only estimate the dark matter mass based on the rotation curve of the galaxy. The observed rotation curve, $v_{\mathrm{total}}$, of a galaxy is the result of the mass distribution within the galaxy's halo, which is made up of contributions from the stellar, gaseous and dark matter components. These can be modelled as
\begin{equation}
	v_{\mathrm{total}}^2(r)=v_{\mathrm{gas}}^2(r)+v_{\mathrm{star}}^2(r)+v_{\mathrm{dm}}^2(r),
	\label{equ:rot_cur}
\end{equation}
where $v_x(r)$ are the radial velocity contributions from the gas, stars and dark matter ($x=$ gas, star, dm; respectively). The gas and stellar surface densities, $\Sigma_{\mathrm{gas}}$ and $\Sigma_{\mathrm{star}}$, are used to estimate the velocity contributions $v_{\mathrm{gas}}$ and $v_{\mathrm{star}}$, respectively.

\subsubsection{Gas Surface Density}
\label{ss-sec:gas_sd}

Under the assumption that H\,\textsc{i} gas is optically thin, the gas surface density of a galaxy is related to its H\,\textsc{i} column density profile using,
\begin{equation}
	\Sigma_{\mathrm{gas}}(r)=fm_{\mathrm{H}}N_{\mathrm{HI}}(r)\cos(i),
	\label{equ:gas_sd}
\end{equation}
where $f$ is a scale factor used to account for the contributions from helium and molecular gas \citep[we set $f=1.4$, as often assumed in the literature, e.g.][]{Westmeier2011,Reynolds2019,Elagali2019}, $m_{\mathrm{H}}$ is the mass of the hydrogen atom, $N_{\mathrm{HI}}(r)$ is the H\,\textsc{i} column density and $i$ is the inclination of the galaxy. We extract the H\,\textsc{i} surface density using the \textsc{gipsy} \citep{vanderHulst1992,Vogelaar2001} task \textsc{ellint} for rings with a width of 15\,arcsec defined by the inclination and position angle that we used to extract the rotation curve from the tilted ring modelling (see Table~\ref{table:fit_params}). Additionally, we assume that the gas is located within an infinitely thin disc. If there are regions containing dense gas that is not optically thin then $\Sigma_{\mathrm{gas}}$ will only provide a lower limit at these locations. The gas surface density is shown by the blue squares in the upper panel of Figure~\ref{fig:mass_model}. In Table~\ref{table:fit_params}, gas surface densities, $\Sigma_{\mathrm{gas}}$, measured at radii $>82.5$\,arcsec correspond to the H\,\textsc{i} tail, which is not fit by the tilted ring modelling.

\subsubsection{Stellar Surface Density}
\label{ss-sec:star_sd}

Stellar surface densities, $\Sigma_{\mathrm{star}}$, are derived from optical or near-infrared flux densities using wavelength-dependent stellar mass-to-light ratios. Here we derive $\Sigma_{\mathrm{star}}$ using VISTA Hemisphere Survey \citep[VHS,][]{McMahon2013} $J$- and $K$-band images. Near-infrared data have the advantage over optical data of being less sensitive to absorption by dust \citep[e.g.][]{Taylor2011,Westmeier2011}. We use equations~6 and 7 from \cite{Reynolds2019} to convert the $J$- and $K$-band image pixel units, $A_{J}$ and $A_{K}$, to maps of stellar mass. Similar to the H\,\textsc{i} surface density described above, we again use \textsc{ellint} to extract the stellar surface density in 5\,arcsec rings with the parameters used in the tilted ring model (see Table~\ref{table:fit_params}). We use 5\,arcsec rings instead of 15\,arcsec rings due to the high angular resolution of the VHS images ($\sim0.51$\,arcsec). We then take the average of the surface densities derived from the $J$- and $K$-band images and interpolate between rings to determine the $\Sigma_{\mathrm{star}}$ at the same radii as $\Sigma_{\mathrm{gas}}$ and the derived rotation curve (Table~\ref{table:fit_params}). The average stellar surface density derived using 5\,arcsec and interpolated to 15\,arcsec rings are shown by the hollow grey and filled orange circles, respectively, in the upper panel of Figure~\ref{fig:mass_model}. We note that if we instead extract $\Sigma_{\mathrm{star}}$ using 15\,arcsec rings our results do not change within the stellar surface density uncertainties.

Unlike the gas, which we assume to be in an infinitely thin disc, we assume that the stars are in a disc with non-negligible height. We model vertical density of the stellar disc assuming it has a $\mathrm{sech}^2(z/z_0)$ dependence given by,
\begin{equation}
	\rho(r,z)=\rho(r)\mathrm{sech}^2(z/z_0),
	\label{equ:vertical_density}
\end{equation}
where $\rho(r)$ is the radial density distribution, $z$ is the height above the disc and $z_0$ is the disc scale height. This vertical density dependence is based on observations of edge-on spiral galaxies \citep{vanderkruit1981a,vanderkruit1981b} and is frequently used in the literature \citep[e.g.][]{deblok2008,Westmeier2011,Reynolds2019,Elagali2019}. We assume a ratio between the stellar disc exponential scale length, $l$, and scale height of $l/z_0=5$ \citep[e.g.][]{vanderkruit1981a,Westmeier2011}. Using the average stellar surface density profile, we calculate the stellar disc scale length to be $l=7.9\pm0.9$\,kpc, which gives a disc scale height of $z_0=1.6\pm0.9$\,kpc.

\subsubsection{Dark Matter Profile}
\label{ss-sec:dark_matter}

There are a number of profiles that can be used to model the dark matter density distribution, such as pseudo-isothermal \citep{Begeman1991}, Burkert (\citeyear{Burkert1995}) and Navarro, Frenk and White \citep[NFW,][]{Navarro1997}. Here we have chosen to use the NFW profile, which is based on the results of N-body simulations. We limit our analysis to using only the NFW profile due to the low spatial resolution (i.e.\ $\sim5$\,kpc) of the H\,\textsc{i} data. We would require higher spatial resolution of the inner $\sim5$--10\,kpc to assess the goodness of fit of different dark matter profiles. For details of the NFW profile see section~4.3.3 of \cite{Westmeier2011} and their equations~18 and 19 for the NFW density profile and circular velocity profile that we use in this work. 

\subsubsection{Dark Matter Mass}
\label{ss-sec:dm_mass}

\begin{figure}
	\includegraphics[width=\columnwidth]{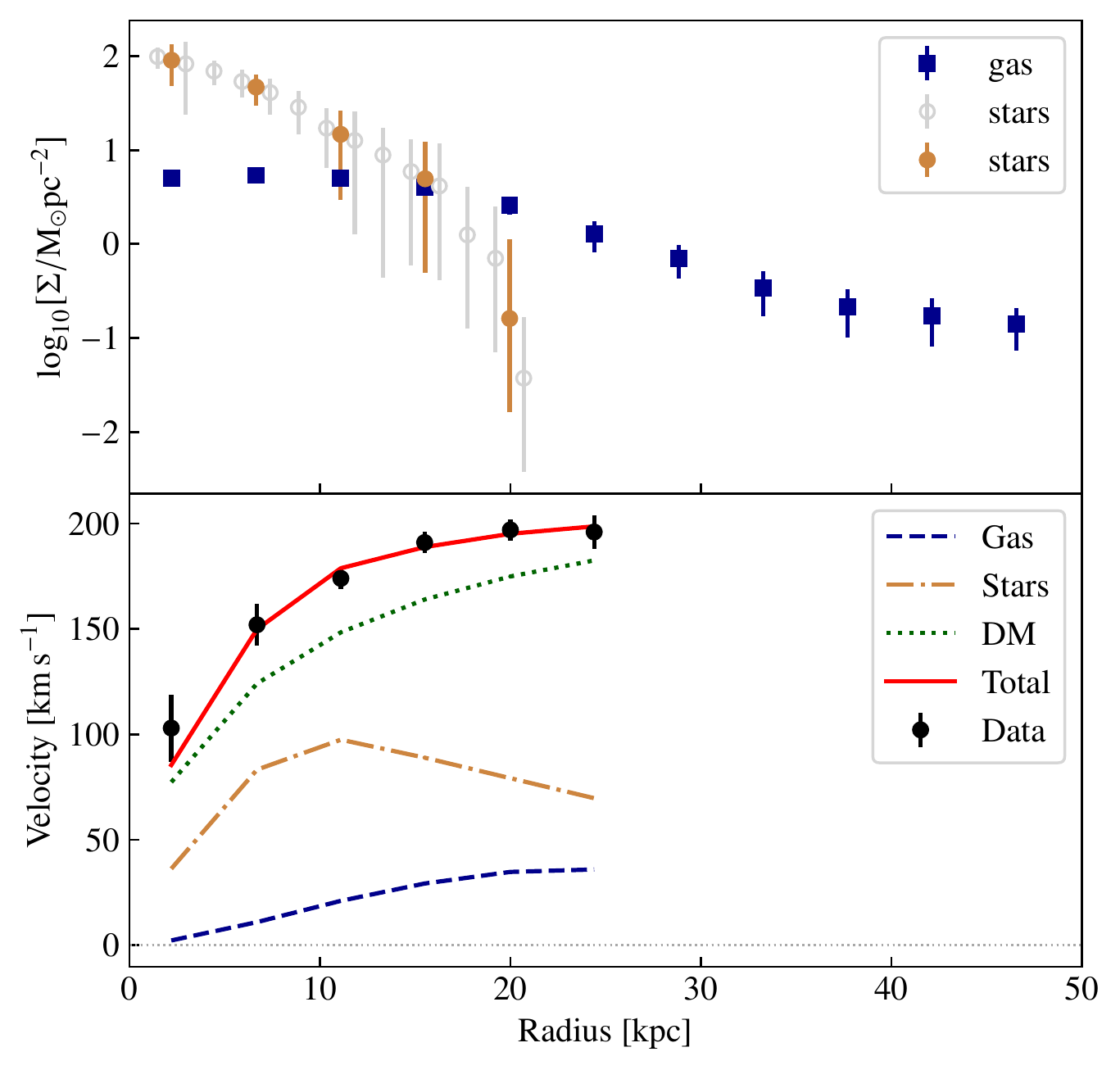}
    \caption{The upper panel shows the gas and average stellar surface densities as a function of radius. The stellar surface density is the average from the $J$- and $K$-bands measured using 5\,arcsec rings (grey open circles) and interpolated to the H\,\textsc{i} resolution of 15\,arcsec rings (filled orange circles). The gas surface density is measured from the moment 0 map (filled blue squares). The lower panel shows the results from the mass modelling for an NFW dark matter halo. The blue dashed curve, orange dot-dashed curve and green dotted curve show the contributions from the gas, stars and dark matter to the total rotation curve (red solid curve) which is plotted over the observed rotation curve (black circles).}
    \label{fig:mass_model}
\end{figure}

To determine the dark matter mass required for the observed rotation curve we use the \textsc{gipsy} task \textsc{rotmas}. The inputs for \textsc{rotmas} are the observed rotation curve (derived from the tilted ring modelling of the gas), the stellar and gaseous surface densities (derived from the VHS $J$- and $K$-band images and H\,\textsc{i} column density map, respectively) and the velocity profile for our chosen dark matter density profile (NFW). We show our derived mass model in the lower panel of Figure~\ref{fig:mass_model}, which shows good agreement between the observed rotation curve (black circles) and derived total rotation curve (solid red curve). The individual velocity contributions from stars, gas and dark matter to the overall rotation curve are shown by the dot-dashed (orange), dashed (blue) and dotted (green) curves, respectively. We tabulate the derived fit parameters, total dark matter mass and dark matter fraction in Table~\ref{table:dm_model_param}. 

We estimate the dark matter mass within a radius of $r<24.4$\,kpc (the largest radius at which we have derived the rotational velocity) to be $\log(M_{\mathrm{DM}}/[\mathrm{M}_{\sun}])=11.3\pm0.1$, which results in a dark matter fraction of $f_{\mathrm{DM}}=0.86$ indicating that ESO\,501$-$G075 is dark matter dominated. Our total mass (gas $+$ stars $+$ dark matter) derived from the mass modelling of $\log(M_{\mathrm{total}}/[\mathrm{M}_{\sun}])=11.3\pm0.1$ is also in agreement with the dynamical mass estimate of $\log(M_{\mathrm{Dyn}}/[\mathrm{M}_{\sun}])=11.3\pm0.1$ within $r<24.4$\,kpc. 

Placing our dark matter mass estimate in context, we compare the dark matter fraction for ESO\,501$-$G075 within $<2.2l=15.9$\,kpc ($l$ is the disc scale length) to the results for disc galaxies from \cite{Courteau2015}. We estimate $f_{\mathrm{DM}}=0.78$ for $r<15.5$\,kpc ($v_{\mathrm{rot}}\sim191$\,km\,s$^{-1}$). This is in agreement with the DiskMass Survey \citep{Bershady2010,Martinsson2013} disc galaxies in figure~1 from \cite{Courteau2015} and show ESO\,501$-$G075 has a dark matter fraction and rotational velocity consistent with possessing stable stellar and gaseous discs.

\begin{table}
	\centering
    \caption{Dark matter parameters output from the mass modelling: $r_{\mathrm{s}}$ is the dark matter profile scale radius, $r_{200}$ is the radius at which the mean halo density is 200 times $\rho_{\mathrm{crit}}$, $\chi^2_{\mathrm{red}}$ is the reduced $\chi^2$ for the mass model fit, $\log(M_{\mathrm{DM}}/[\mathrm{M}_{\sun}])$ is the dark matter mass within $r<24.4$\,kpc and $f_{\mathrm{DM}}$ is the dark matter fraction.}
	\label{table:dm_model_param}
	\begin{tabular}{lr}
		\hline
		Parameter & Value \\
        \hline
		$r_{\mathrm{s}}$ [kpc]                      & $32\pm7$ \\
		$r_{200}$ [kpc]                             & $141\pm14$ \\
		$\chi^2_{\mathrm{red}}$                     & 0.7 \\
		$\log(M_{\mathrm{DM}}/[\mathrm{M}_{\sun}])$ & $11.3\pm0.1$ \\
		$f_{\mathrm{DM}}$                           & 0.86 \\
		\hline
	\end{tabular}
\end{table}

\section{Discussion}
\label{sec:discussion}

We now address whether ESO\,501$-$G075's observed H\,\textsc{i} morphology is compatible with it infalling into the Hydra I cluster and with the H\,\textsc{i} disc being disturbed by environmental processes.

As mentioned in the introduction, ram pressure stripping is thought to be the dominant mechanism that removes H\,\textsc{i} from cluster galaxies. Although we note that the absence of significant H\,\textsc{i} stripping (i.e.\ average H\,\textsc{i} content $\mathrm{DEF}_{\mathrm{HI}}=0.14$) indicates that ESO\,501$-$G075 is likely infalling for the first time and any mechanism acting on it has yet to remove significant quantities of H\,\textsc{i} gas from the galaxy. This is also supported by ESO\,501$-$G075's position in projected phase-space (Figure~\ref{fig:phase_space}), which provides a minimum distance from the cluster centre ($\sim1.05^{\circ}$).

\begin{figure}
	\includegraphics[width=\columnwidth]{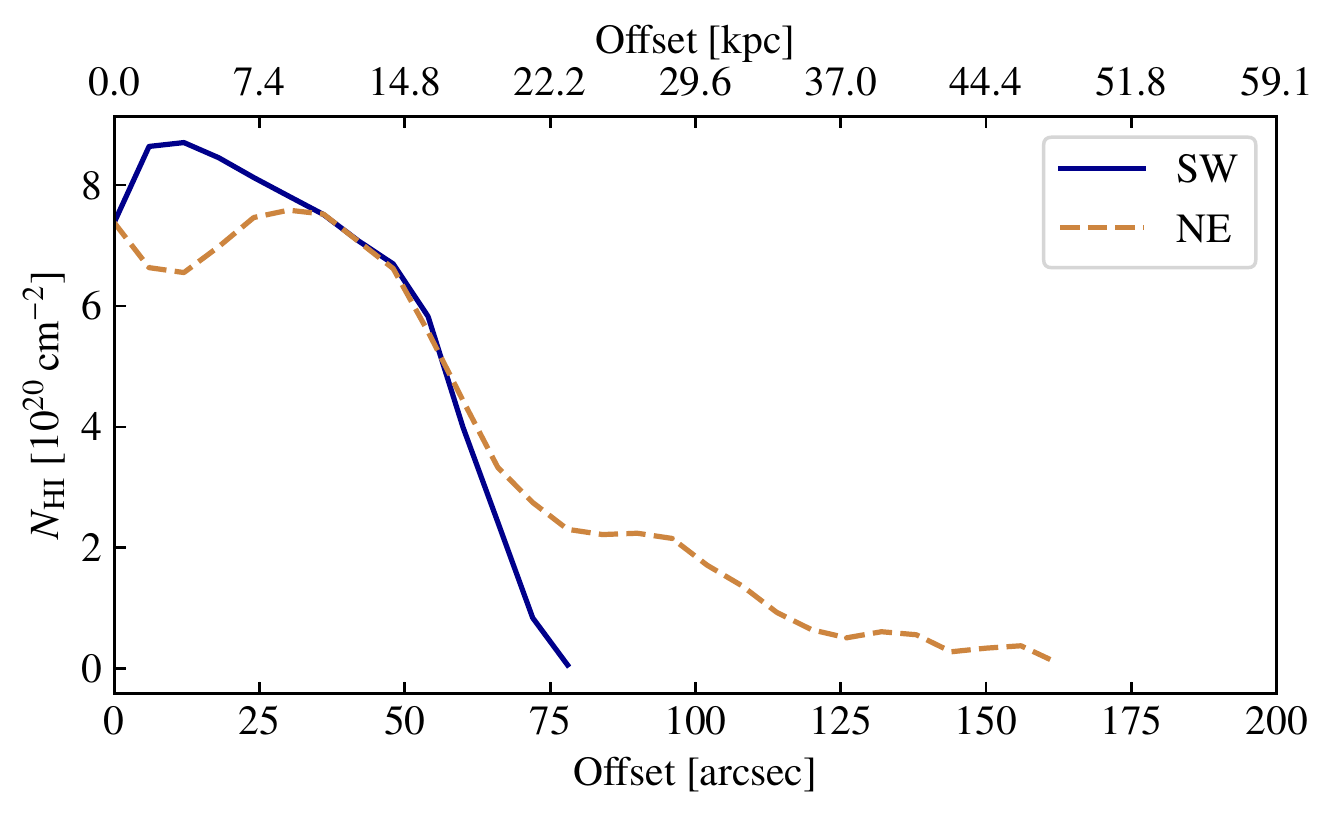}
    \caption{Column density cut along the major axis of ESO\,501$-$G075, extracted from the 6\,arcsec wide red dashed box indicated in Figure~\ref{fig:moment0_map} illustrating that the H\,\textsc{i} disc is more extended on the NE (orange) compared to the SW (blue) side.}
    \label{fig:colden_cut}
\end{figure}

In Figure~\ref{fig:colden_cut}, we show the column density profile taken as a cut through the centre of ESO\,501$-$G075 along the major axis. This illustrates the difference in the H\,\textsc{i} distribution between the two sides of the disc (i.e.\ compression of H\,\textsc{i} column density contours to the SW and elongation of contours to the NE of the galaxy centre, seen in Figure~\ref{fig:moment0_map}). Figure~\ref{fig:colden_cut} shows the sharp, rapid drop in column density on the SW (blue) side compared with the more gradual and extended decline in column density on the NE (orange) side at distances $\gtrsim75$\,arcsec. The absence of close neighbouring galaxies (Figure~\ref{fig:sky_map}, Table~\ref{table:all_galaxies}) and the orientation of the H\,\textsc{i} extended tail pointing away from the cluster centre suggest that the observed morphology may be a result of ESO\,501$-$G075 interacting with the cluster environment.

We investigate whether the observed H\,\textsc{i} morphology is consistent with a ram pressure stripping scenario using a simple model that equates the strength of the ram pressure exerted on the gaseous disc to the gravitational restoring force produced by the dark matter halo of ESO\,501$-$G075 as a function of disc radius. \cite{Gunn1972} proposed this method for estimating the radius beyond which gas can be removed due to ram pressure stripping (e.g.\ the stripping radius) from a galaxy travelling face-on through the IGM. The gravitational restoring force cannot be estimated at a height above the disc if the galaxy is instead moving edge-on through the IGM. However, for galaxy inclinations with respect to the infall direction $\lesssim60^{\circ}$ (not to be confused with the observed inclination, $i$), \cite{Roediger2006} show that this method for equating the gravitational restoring force to ram pressure produces meaningful results in simulations.

Assuming that a galaxy is moving face-on through the cluster IGM, the only gravitational restoring force acting on the galaxy is perpendicular to the disc (i.e.\ in the $z$-direction). In this scenario, the gravitational restoring force at a given radius, $r$, within the disc of a galaxy can be estimated from \citep[e.g.][]{Roediger2006,Koppen2018},
\begin{equation}
	P_{\mathrm{grav}}(r)=\Sigma_{\mathrm{gas}}(r)\left|\frac{\partial \Phi(r)}{\partial z}\right|_{\mathrm{max}},
	\label{equ:p_grav}
\end{equation}
where $\Sigma_{\mathrm{gas}}(r)$ is the gas surface density and $\left|\frac{\partial \Phi(r)}{\partial z}\right|_{\mathrm{max}}$ is the gravitational acceleration due to a given gravitational potential $\Phi(r)$. The gravitational acceleration is determined from the height above the disc, $z$, that maximises the acceleration. For an NFW profile the gravitational potential is given by 
\begin{equation}
	\Phi_{\mathrm{NFW}}(r)=\frac{-G M_{200} \ln\left(1+\frac{r_{\mathrm{NFW}}}{r_{\mathrm{s}}}\right)}{r_{\mathrm{NFW}}\left[\ln(1+c)-\frac{c}{1+c}\right]},
	\label{equ:nfw_potential}
\end{equation}
where $r_{\mathrm{NFW}}$ is the radius of the spherical dark matter halo (not to be confused with the radius of the stellar/gaseous disc), $G$ is the gravitational constant, $c=R_{200}/r_{\mathrm{s}}$ is the concentration parameter, $r_{\mathrm{s}}$ is the halo scale radius, $R_{200}$ is the radius at which the mean halo density is 200 times the critical density of the Universe and is approximately the halo virial radius and $M_{200}$ is the mass enclosed within $R_{200}$. $M_{200}$ can be calculated from $R_{200}$ using
\begin{equation}
	M_{200} = \frac{100 H_0^2 R_{200}^3}{G}.
	\label{equ:m200}
\end{equation}

The strength of the ram pressure experienced by the disc is determined by
\begin{equation}
	P_{\mathrm{ram}}=n_{\mathrm{IGM}} v_{\mathrm{rel}}^2,
	\label{equ:p_ram}
\end{equation}
where $n_{\mathrm{IGM}}$ is the density of the IGM and $v_{\mathrm{rel}}$ is the velocity of the galaxy relative to the IGM. Ram pressure will dominate and be effective at stripping gas when $P_{\mathrm{grav}}(r)<P_{\mathrm{ram}}$ (e.g.\ below the horizontal dashed lines in Figure~\ref{fig:balance_forces}). The velocity difference between ESO\,501$-$G075 and the Hydra I cluster of $\Delta$c$z_{\mathrm{cl}}\sim960$\,km\,s$^{-1}$ provides an estimate for the relative velocity, $v_{\mathrm{rel}}$. We estimate the ram pressure strength, $P_{\mathrm{ram}}$, for $v_{\mathrm{rel}}=800$, 1\,000 and 1\,200\,km\,s$^{-1}$.

\begin{figure}
	\includegraphics[width=\columnwidth]{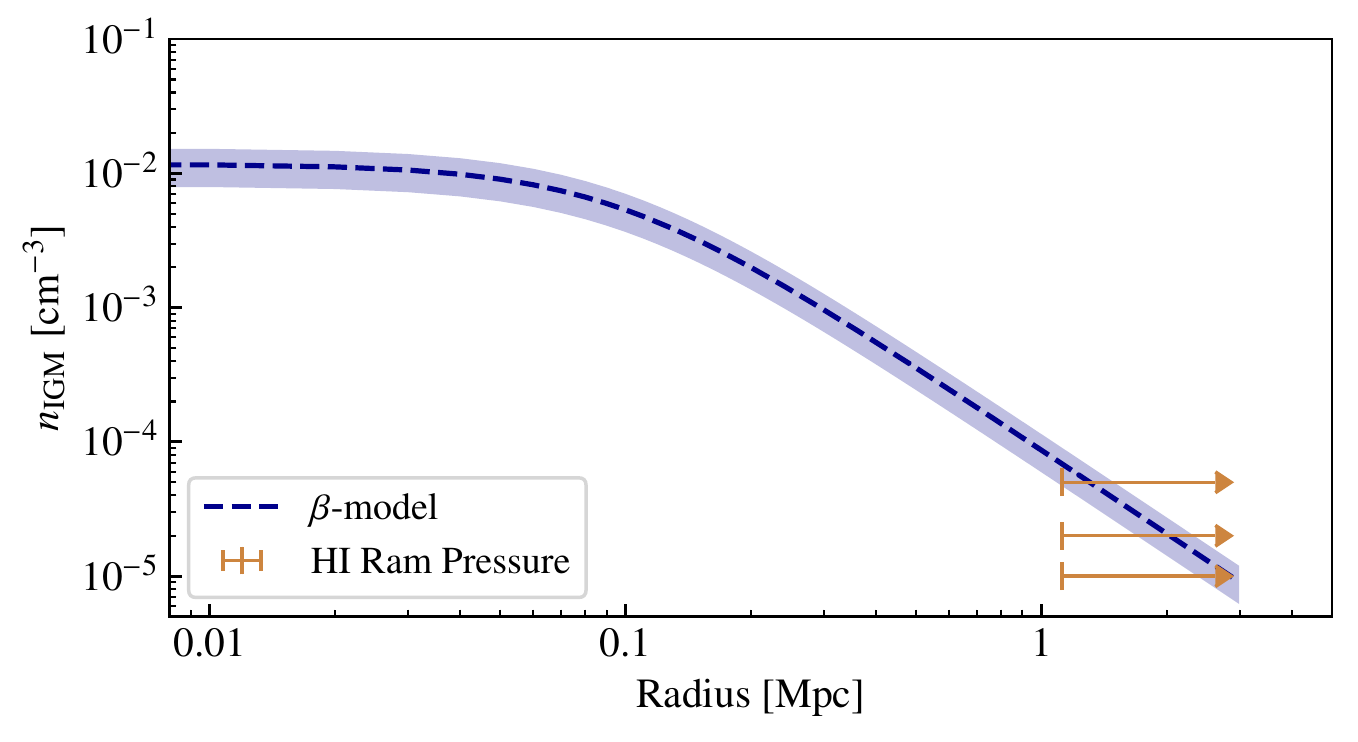}
    \caption{Hydra I cluster IGM $\beta$-model density profile. The dashed blue curve shows the profile for $n_0=0.011$\,cm$^{-3}$ with the shaded region indicating the variation for central densities of $\pm0.003$\,cm$^{-3}$ \citep[i.e.\ the difference in central density from \textit{Chandra} and XMM-Newton observations,][]{Hayakawa2004,Hayakawa2006}. The orange lines indicate the three IGM densities, $n_{\mathrm{IGM}}$, we use for estimating the strength of the ram pressure exerted on ESO\,501$-$G075 (see Figure~\ref{fig:balance_forces}) with the vertical line located at the projected distance of ESO\,501$-$G075 from the cluster centre and the arrow indicating that this is a lower limit on the 3-dimensional cluster centric distance.}
    \label{fig:density_profile}
\end{figure}

The optimal method for estimating the IGM density is using X-ray observations. However, the Hydra I cluster only has \textit{Chandra} and XMM-Newton X-ray observations out to $\sim50$\,kpc \citep{Hayakawa2004, Hayakawa2006, Cavagnolo2009}. Instead, we assume a $\beta$-model \citep[e.g.][]{Cavaliere1976,Gorenstein1978} to model the radial cluster density profile, which is given by,
\begin{equation}
	n_{\mathrm{IGM}}(r) = n_0 \left(1+\left(\frac{r}{r_{\mathrm{c}}}\right)^2 \right)^{-3\beta/2},
	\label{equ:beta_model}
\end{equation}
where $n_0$ is the central density of the cluster, $r_{\mathrm{c}}$ is the core radius and $\beta$ characterises the radial density profile. We set $n_0=0.011$\,cm$^{-3}$, $r_{\mathrm{c}}=94$\,kpc and $\beta=0.69$ based on XMM-Newton X-ray observations of the Hydra I cluster from \citep{Hayakawa2006}. Using these values we derive the density profile shown in Figure~\ref{fig:density_profile}. We note the $\beta$-model assumes that the IGM density follows an isotropic, perfectly smooth distribution. This is unlikely to be a true representation of the local IGM density at ESO\,501$-$G075 as clusters are frequently found to contain significant density structure \cite[e.g.][]{Jones1999, Schuecker2005, Parekh2015}. We assume IGM densities of $n_{\mathrm{IGM}}=1$, 2 and $5\times10^{-5}$\,cm$^{-3}$ for estimating $P_{\mathrm{ram}}$ as the $\beta$-model IGM density is $\lesssim5\times10^{-5}$\,cm$^{-3}$ at distances $\geq1.12$\,Mpc (i.e.\ the projected cluster centric distance). The projected distance provides a lower limit on the true, 3-dimensional distance with the galaxy likely $>1.12$\,Mpc based on its velocity relative to the cluster and position in phase-space (Figure~\ref{fig:phase_space}).

\begin{figure}
	\includegraphics[width=\columnwidth]{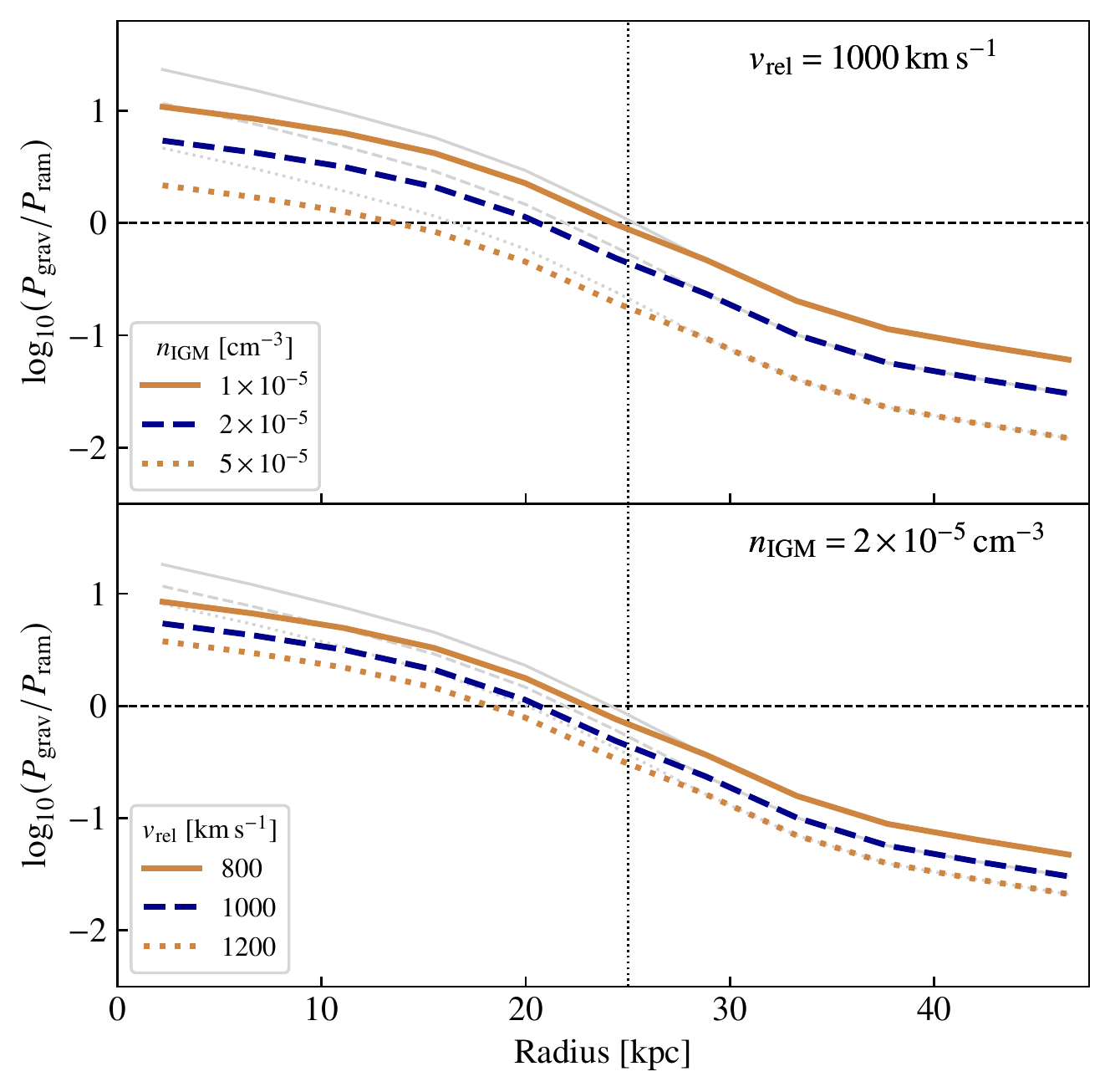}
    \caption{The ratio of the gravitational restoring pressure to the ram pressure felt by gas at a given radius from the galaxy centre. The upper panel shows the variation in the pressure ratio for three assumed IGM densities, $n_{\mathrm{IGM}}$, and a fixed relative velocity of the galaxy with respect to the IGM of $v_{\mathrm{rel}}=1\,000$\,km\,s$^{-1}$. The lower panel show the same but for three assumed relative velocities, $v_{\mathrm{rel}}$, and a fixed IGM density of $n_{\mathrm{IGM}}=2\times10^{-5}$\,cm$^{-3}$. Ram pressure dominates at radii where the curves drop below the horizontal dashed line. The vertical dotted line indicates the approximate radius of the presumed leading edge of the H\,\textsc{i} disc (SW side of galaxy in the left panel of Figure~\ref{fig:moment0_map}). The grey curves indicate the effect of including the gravitational potential for the stellar disc in $P_{\mathrm{grav}}$.}
    \label{fig:balance_forces}
\end{figure}

In Figure~\ref{fig:balance_forces}, we show the radial variation in the ratio of the gravitational restoring force to the ram pressure strength for three values of the IGM density, $n_{\mathrm{IGM}}$, and three galaxy velocities relative to the IGM, $v_{\mathrm{rel}}$. In the upper panel we fix $v_{\mathrm{rel}}=1\,000$\,km\,s$^{-1}$ and vary $n_{\mathrm{IGM}}$ while in the lower panel we fix $n_{\mathrm{IGM}}=2\times10^{-5}$\,cm$^{-3}$ and vary $v_{\mathrm{rel}}$. For the inner 10--20\,kpc, the gravitational restoring force dominates by up to an order of magnitude indicating ESO\,501$-$G075 is stable against the effects of ram pressure stripping over the stellar disc. In reality, the gravitational restoring force will be even stronger in the inner regions of the galaxy as we have not taken into account the gravitational contributions from the stellar and gaseous discs. 

We illustrate the effect including the stellar disc gravitational potential has on the restoring force, $P_{\mathrm{grav}}$, by modelling the gravitational potential of the stellar disc by the Miyamoto-Nagai potential \citep{Miyamoto1975}, which models a disc with a non-zero thickness and is given by,
\begin{equation}
	\Phi_{\mathrm{MN}}(r,z)=\frac{-G M_{\mathrm{star}}}{\sqrt{r^2 + (\sqrt{z^2 + b^2} + a)^2}},
	\label{equ:disk_potential}
\end{equation}
where $M_{\mathrm{star}}$ is the stellar mass, $r$ is the radius in the plane of the disc, $z$ is the height above the disc and $a$ and $b$ are the disc scale length and height, respectively. The total gravitational potential is then the sum of $\Phi_{\mathrm{NFW}}(r)$ and $\Phi_{\mathrm{MN}}(r,z)$ at radii within the stellar disc ($r<25$\,kpc) and simplifies to $\Phi_{\mathrm{NFW}}(r)$ beyond the stellar disc. The resulting curves for the ratio $\log(P_{\mathrm{grav}}/P_{\mathrm{ram}})$ are shown in grey in Figure~\ref{fig:balance_forces} and are increased by $\sim0.5$\,dex within the stellar disc.

There are some inherent limitations in our assumed simple analytic model for ram pressure stripping. Our assessment of the effectiveness of ram pressure stripping of the H\,\textsc{i} disc assumes that ESO\,501$-$G075 is travelling face-on through the IGM. However, ESO\,501$-$G075 is not likely to be oriented face-on, but instead inclined with respect to the direction of motion through the IGM, which will introduce some uncertainty. In particular, if the galaxy is moving edge-on through the IGM then our simple model of the gravitational restoring force breaks down and does not represent the force that ram pressure must overcome to dominate (i.e.\ our analysis estimates the most extreme amount of stripping that ESO\,501$-$G075 could experience). This explains how our modelling can produce the expectation that gas should be stripped at smaller radii than observed, and indicates that ESO\,501$-$G075 is most likely not moving directly face-on through the IGM (e.g.\ \citeauthor{Jachym2009} \citeyear{Jachym2009} show in simulations that the stripping efficiency decreases as a galaxy's orientation becomes more edge-on, which results in less mass loss and gas being retained at larger radii). However, the model we use can provide a meaningful assessment of the effectiveness of ram pressure for galaxies with orientations $<60^{\circ}$ \citep[e.g.][]{Roediger2006} with respect to the IGM (where face-on is $0^{\circ}$ and edge-on is $90^{\circ}$).

In our model, ram pressure dominates at radii $r>25$\,kpc (the edge of the H\,\textsc{i} disc on the SW side of ESO\,501$-$G075, indicated by the vertical dotted line in Figure~\ref{fig:balance_forces}) for all $n_{\mathrm{IGM}}$ and $v_{\mathrm{rel}}$. Although this is compatible with a ram pressure stripping scenario, which is consistent with our understanding of the cluster environment, there are other environmental mechanisms that could produce the observed morphology of ESO\,501$-$G075. As previously mentioned, tidal interactions, harassment and mergers are unlikely to be responsible, as ESO\,501$-$G075 does not have any close neighbours with or without H\,\textsc{i} detections and the stellar disk appears undisturbed (Section~\ref{s-sec:eso501-g075}). We cannot disentangle the contributions of other hydrodynamical processes (e.g.\ viscous or tidal stripping) from ram pressure. However, viscous stripping is found to act over longer timescales \citep[e.g.\ $\sim10^9$\,yr,][]{Quilis2000,Roediger2005} than ram pressure (e.g.\ $\sim10^7$--$10^8$\,yr, \citeauthor{Abadi1999} \citeyear{Abadi1999}; \citeauthor{Vollmer2001} \citeyear{Vollmer2001}) and tidal stripping should be most effective near the cluster core radius and decrease in strength towards the edge of the cluster \citep{Merritt1984}. 

\section{Conclusions}
\label{sec:conclusion}

We have presented results from the first cluster observed with WALLABY: the Hydra I cluster. These are the first observations reaching the final sensitivity of WALLABY. We provide a first look at Hydra I cluster members within the cluster virial radius detected in H\,\textsc{i} emission and find WALLABY detects 51 galaxies. Of these sources, there are five galaxies spatially resolved in H\,\textsc{i} by $>5$ synthesised beams. We identify two of these galaxies with large asymmetries in their H\,\textsc{i} morphologies. One lies in projection near the cluster centre and shows signs of experiencing ram pressure. The other, ESO\,501$-$G075, lies near the virial radius and has a line of sight velocity relative to the cluster that is near the cluster escape velocity.

We carry out a case study on ESO\,501$-$G075 to assess whether the observed H\,\textsc{i} morphology is compatible with a ram pressure stripping scenario. We derive the rotation curve by fitting a tilted ring model to the H\,\textsc{i} spectral line cube. We then combine the rotation curve with stellar and gaseous radial surface density profiles from averaged VHS $J$- and $K$-band images and the H\,\textsc{i} column density map, respectively, to estimate the dark matter mass of ESO\,501$-$G075 for an assumed NFW dark matter profile. We estimate ESO\,501$-$G075 to have a dark matter mass of $\log(M_{\mathrm{DM}}/[\mathrm{M}_{\sun}])=11.3\pm0.1$ contained within a radius of $r<24.4$\,kpc and find ESO\,501$-$G075 to be dark matter dominated with a dark matter fraction of $f_{\mathrm{DM}}=0.86$. We also find that the dark matter fraction within 2.2 scale lengths ($r<15.9$\,kpc) and rotational velocity ($\sim0.78$ and $\sim191$\,km\,s$^{-1}$, respectively) are consistent with ESO\,501$-$G075 having stable stellar and gaseous discs, consistent with the results of \cite{Courteau2015}.

We then calculate the strength of ram pressure that ESO\,501$-$G075 could experience for IGM densities in the range $n_{\mathrm{IGM}}=1$--$5\times10^{-5}$\,cm$^{-3}$ and relative velocities between ESO\,501$-$G075 and the cluster IGM of $v_{\mathrm{rel}}=800$--1\,200\,km\,s$^{-1}$ (i.e.\ velocities in the range of the velocity difference between the galaxy and the cluster, $\Delta$c$z_{\mathrm{cl}}\sim960$\,km\,s$^{-1}$). We use a simple analytic model to equate the estimated ram pressure to the gravitational restoring force of the galaxy due to an NFW dark matter potential to determine if the observed morphology is consistent with a ram pressure stripping scenario. In our model, ram pressure dominates at radii $r\gtrsim25$\,kpc, which corresponds to the radius of the presumed leading edge of the H\,\textsc{i} disc, for all combinations of IGM density and relative velocity. We conclude that the H\,\textsc{i} morphology of ESO\,501$-$G075 is compatible with the galaxy experiencing ram pressure and consistent with ESO\,501$-$G075 falling into Hydra I for the first time. Our results are also in agreement with the ram pressure study of Hydra I by \cite{Wang2021}, in which ESO\,501$-$G075 is classified as a candidate for experiencing ram pressure.

Looking ahead, WALLABY will spatially resolve $\sim5\,000$ galaxies in H\,\textsc{i} by $>5$ synthesised beams \citep{Koribalski2020}. These galaxies will span a range of environments including groups and clusters. Using this analysis for the effectiveness of ram pressure acting on these galaxies will enable a systematic study of the role that ram pressure plays in removing gas from galaxies in groups and clusters.

\section*{Acknowledgements}

This research was conducted by the Australian Research Council Centre of Excellence for All Sky Astrophysics in 3 Dimensions (ASTRO 3D), through project number CE170100013. We thank the referee for their comments.

The Australian SKA Pathfinder is part of the Australia Telescope National Facility which is managed by the Commonwealth Scientific and Industrial Research Organisation (CSIRO). Operation of ASKAP is funded by the Australian Government with support from the National Collaborative Research Infrastructure Strategy. ASKAP uses the resources of the Pawsey Supercomputing Centre. Establishment of ASKAP, the Murchison Radio-astronomy Observatory (MRO) and the Pawsey Supercomputing Centre are initiatives of the Australian Government, with support from the Government of Western Australia and the Science and Industry Endowment Fund. We acknowledge the Wajarri Yamatji as the traditional owners of the Observatory site. We also thank the MRO site staff. This paper includes archived data obtained through the CSIRO ASKAP Science Data Archive, CASDA (\url{http://data.csiro.au}). 

The Parkes radio telescope is part of the Australia Telescope National Facility which is funded by the Commonwealth of Australia for operation as a National Facility managed by CSIRO.

This project has received funding from the European Research Council (ERC) under the European Union’s Horizon 2020 research and innovation programme (grant agreement no. 679627; project name FORNAX). 

This project has received support from the BMBF project 05A17PC2 for D-MeerKAT.

SHO acknowledges support from the National Research Foundation of Korea (NRF) grant funded by the Korea government (Ministry of Science and ICT: MSIT) (No. NRF-2020R1A2C1008706).

ARHS acknowledges receipt of the Jim Buckee Fellowship at ICRAR-UWA.

LC is the recipient of an Australian Research Council Future Fellowship (FT180100066) funded by the Australian Government.

This research has made use of the NASA/IPAC Extragalactic Database (NED) which is operated by the Jet Propulsion Laboratory, California Institute of Technology, under contract with the National Aeronautics and Space Administration.

This work is based in part on observations made with the Galaxy Evolution Explorer (\textit{GALEX}). \textit{GALEX} is a NASA Small Explorer, whose mission was developed in cooperation with the Centre National d'Etudes Spatiales (CNES) of France and the Korean Ministry of Science and Technology. \textit{GALEX} is operated for NASA by the California Institute of Technology under NASA contract NAS5-98034.

\section*{Data Availability}

The full 36 beam, 30-square-degree H\,\textsc{i} spectral line cubes are available from the CSIRO ASKAP Science Data Archive \citep[CASDA,][]{Chapman2015,Huynh2020} using the DOI \url{https://doi.org/10.25919/5f7bde37c20b5}.




\bibliographystyle{mnras}
\bibliography{master} 







\bsp	
\label{lastpage}
\end{document}